\documentclass[groupedaddress, reprint,superscriptaddress,amsmath,amssymb,aps,pra]{revtex4-2}
\usepackage{comment}
\usepackage{graphicx}% Include figure files
\usepackage{dcolumn}% Align table columns on decimal point
\usepackage{bm}% bold math
\usepackage{hyperref}% add hypertext capabilities
\usepackage{cleveref} % for improved referencing
\usepackage{xcolor}
\usepackage{soul}
\usepackage{physics}
\usepackage{upgreek}

\usepackage{bbm}

\usepackage{tensor}

\usepackage[normalem]{ulem}
\usepackage{mathtools}
\begin{document}
\preprint{APS/123-QED}

\title{Optimal single-mode squeezing for beam displacement sensing}

\author{Wenhua He}
\affiliation{Wyant College of Optical Sciences, University of Arizona, Tucson, AZ 85721, USA}
\author{Christos N. Gagatsos}
\affiliation{Department of Electrical and Computer Engineering, University of Arizona, Tucson, AZ 85721, USA}
\affiliation{Wyant College of Optical Sciences, University of Arizona, Tucson, AZ 85721, USA}
\affiliation{Program in Applied Mathematics, University of Arizona, Tucson, AZ 85721, USA}
\author{Dalziel J. Wilson}%
\affiliation{Wyant College of Optical Sciences, University of Arizona, Tucson, AZ 85721, USA}
\author{Saikat Guha}%
\affiliation{Department of Electrical and Computer Engineering, University of Maryland, College Park, MD 20742, USA}
\affiliation{Wyant College of Optical Sciences, University of Arizona, Tucson, AZ 85721, USA}

\begin{abstract}
    Estimation of an optical beam's transverse displacement is a canonical imaging problem fundamental to numerous optical imaging and sensing tasks. Quantum enhancements to the measurement precision in this problem have been studied extensively. However, previous studies have neither accounted for diffraction loss in full generality, nor have they addressed how to jointly optimize the spatial mode and the balance between squeezing and coherent amplitude. Here we show that, in the small-displacement limit, the seemingly intractable infinite-spatial-mode problem can be reduced to a compact three-mode interaction framework. We quantify the improvement afforded by an optimized single-spatial-mode Gaussian-state probe over the optimal classical laser probe, and show that a two-spatial-mode homodyne receiver is asymptotically optimal for the former in the limit of high probe energy. Our findings reveal a strategy for identifying quantum-optimal probes in the presence of generic multimode linear probe-target interaction and photon loss.
    \end{abstract}

\maketitle

\section{Introduction}
Quadrature squeezed states of light have been at the center stage of quantum-enhanced photonic sensors~\cite{schnabel2017squeezed,andersen2016squeezedlightgeneration,lawrie2019squeezedlightsensingreview}, most notably at the Laser Interferometer Gravitational-Wave Observatory (LIGO)~\cite{ganapathy2023broadbandfreq_squeezing}. The enhanced sensitivity is afforded by the {\em squeezed} quadrature noise in a coherent-detection measurement such as homodyne detection~\cite{schnabel2017squeezed,lawrie2019squeezedlightsensingreview,andersen2016squeezedlightgeneration}, which results in a higher-precision estimation of an unknown optical phase ~\cite{lang2013optimal,lang2014optimal,monras2006optimal}. In many scenarios, the interaction with the target, of $M$ orthogonal spatio-temporal-polarization modes of an optical probe, imparts correlated single-mode phases on the constituent modes---expressible as a passive linear lossy $M$-mode interferometer---which collectively encode the single (or a few) parameter(s) of interest of the target. In such situations, a probe that is entangled across the target-modulated modes can yield a large sensitivity boost in estimating the parameter(s) of interest~\cite{Rubio2020,zhuang2018distributed, ge2018distributed}. For a lossless, noiseless probe-target interaction, the mean-squared error (MSE) of estimation goes from the Standard Quantum Limit (SQL) scaling of $\sim 1/M$ (with no entanglement) to the improved Heisenberg Limit (HL) scaling of $\sim 1/M^2$ (with entanglement)~\cite{quantummetrology}. The actual improvement in a realistic setting depends upon many factors, including loss and noise in the probe, the physics underlying the optical probe's interaction with the target~\cite{Xia2020, Grace2020, xia2023entanglement,qi2018ultimate}, and the receiver measurement. 

Squeezed light, when split in a multimode linear interferometer, produces a multimode entangled Gaussian state that can reap the aforesaid entanglement-assisted sensitivity boost~\cite{zhuang2018distributed}. This principle has been demonstrated in the contexts of radio-frequency photonic sensors~\cite{Xia2020, grace2020entanglement}, and opto-mechanical force sensing~\cite{xia2023entanglement}, and has been theorized to reap improved performance in other contexts such as fiber-optical laser gyroscopes~\cite{Mehmet2010,Grace2020}, joint range-velocity estimation for a quantum LAser Detection And Ranging (LADAR) system~\cite{reichert2024heisenberg}, and near-field beam deflection measurements~\cite{qi2018ultimate}. Because of this property of squeezed light, asking which (potentially high-order) mode of the probe field to excite in a squeezed state to achieve the best sensing precision, is also tantamount to optimizing within a subclass of multimode Gaussian {\em entangled} states of the probe field---thereby allowing one to access the HL sensitivity with a {\em single-mode} squeezed state of an appropriate mode. Recent years have seen rapid maturation of techniques for multimode transformations and structuring of both spatial and temporal-spectral modes of squeezed light~\cite{squeezed_light_pump,squeezed-light-cam,La-Volpe2020, kues2019quantumopticalcomb}, which makes it timely to pursue the aforesaid genre of quantum enhanced sensing with multimode and high-order squeezed-light probes. Some experimental studies along these lines include quantum-enhanced laser beam pointing using probe fields excited in tensor products of coherent state of a mode and squeezed state(s) of orthogonal mode(s)~\cite{2003-SD-EXP2D, li2022higher}, and a spectral-mode equivalent of the idea demonstrated in the context of frequency-dependent squeezing-enhanced gravitational wave detection~\cite{ganapathy2023broadbandfreq_squeezing}. 

\begin{figure}
    \centering
\includegraphics[width=0.9\linewidth]{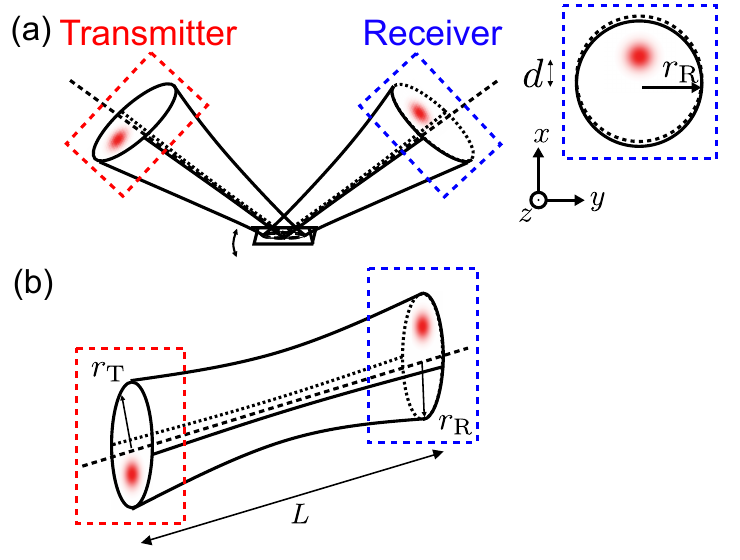}
\vspace{-10pt}
    \caption{(a) Transceiver model for optical deflectometry. (b) Free-space propagation through soft-Gaussian apertures with radius $r_\mathrm{T}$ and $r_\mathrm{R}$ (for our analysis, $r_\mathrm{T}=r_\mathrm{R}=r_0$), separated by $L$. The beam is displaced along the $x$-axis. 
    }
    \label{fig:1}
    \vspace{-10pt}
\end{figure}
High-precision estimation of a transverse displacement of an optical beam~\cite{2000-SD-theory} underlies a wide variety of applications, such as scanning-probe atomic-force microscopy~\cite{putman1992afm}, and pointing, acquisition and tracking in free-space laser communications~\cite{kaymak2018FSOsurvey}. Many classical and quantum protocols have been proposed, both for the transmitter (optical probe) and the receiver (measurement of the target-return light)~\cite{2002-SD-EXP1D, TEM00displacement, 2014-QFI-HGn0}. A pioneering quantum solution, dubbed the {\em quantum laser pointer}, employed a three-mode probe state where two modes were squeezed and one was a coherent state, and the measurement was performed with a quadrant detector~\cite{2003-SD-EXP2D}. Despite a clear quantum advantage, it wasn't clear if that was the best quantum probe, even among all multi-spatial-mode Gaussian-state probes, i.e., states of light accessible with squeezing and linear optics. 

An important theoretical breakthrough that allowed for a formal probe-state optimization was a normal-mode decomposition of the return-path free-space propagation of the probe and target-return light, contained in a near-field free-space propagation volume, book-ended by Gaussian soft-aperture pupils, as a multi-mode linear transformation involving pairwise beamsplitter interferences between \textit{adjacent} Hermite-Gauss (HG) modes (which are the {\em normal modes}~\cite{shapiro2005ultimate} of the above said soft-aperture propagation kernel), but with an omission of the consideration of propagation loss~\cite{qi2018ultimate}. A more general theory of the mode parameter-encoding Hamiltonian and a Quantum-Fisher Information (QFI) assisted analysis was developed first in the context of beam displacement~\cite{He2022} and more recently in a general setting~\cite{2023mode_parameter_estimation}. MSE scales roughly as the inverse of the QFI~\cite{braunstein1994QFI}. When the parameter-encoding transformation is a multimode passive unitary, under a mean-probe-energy constraint, the QFI-optimal Gaussian-state probe was proven to be a single-mode squeezed-vacuum state excited in an appropriate mode with the optimum receiver being a single-mode homodyne detection (in the same mode)~\cite{matsubara2019optimal}. The QFI-optimal probe under a mean energy constraint, even restricted to Gaussian-state probes, when the multimode target interaction is passive but {\em lossy}, remains open. The QFI-optimal spatial mode to excite a classical (laser-light) probe was found to be a bi-modal-shaped mode, which significantly outperforms a laser-light probe in the $\mathrm{HG}_{00}$ mode~\cite{he2024optimum}.

In this paper, inspired by recently-demonstrated experimental feasibility of linear-spatial-mode structuring of squeezed light~\cite{squeezed-light-cam} and generation of squeezing in a high-order spatial mode~\cite{squeezed_light_pump}, as well as the single-mode squeezed-vacuum state's QFI optimality for estimation of a single parameter embedded in a mutlimode passive unitary transformation in the lossless limit~\cite{matsubara2019optimal}, we investigate the optimum spatial mode to excite a---potentially, phase-space-displaced---squeezed-light probe in, and an associated practical receiver design, for transverse displacement (beam-deflection) estimation; and quantify the improvement over optimal spatially-structured classical illumination~\cite{he2024optimum}.

\subsection{Main results}
Our primary mathematical innovation is a representation of the general multimode parameter-encoding Hamiltonian---for {\em any} single-spatial-mode optical probe excited in {\em any} quantum state, in the regime of a small angular beam-deflection---as a three-mode lossy passive interferometer. This decomposition allows for a numerically-tractable optimization of the QFI-optimal single-mode squeezed-state probe, in contrast with previous work~\cite{qi2018ultimate, 2023mode_parameter_estimation}. The decomposition involves: (1) the probe's excited spatial mode, (2) an environment mode in the vacuum state capturing the return-path transmissivity, and (3) an `adjacent' spatial mode into which the probe's photon energy `spills over'. Using this framework, under a mean photon number constraint in the transmitted probe light, and in the near-field diffraction limited regime, we: (1) numerically evaluate the optimal mode in which to excite a squeezed state and the optimum split of the total probe photon energy into the coherent and squeezed contributions, (2) quantify the QFI improvement achieved by this optimum single-mode squeezed-light probe over the optimal laser-light probe found in~\cite{he2024optimum}, and (3) numerically find a receiver design involving a spatial mode demultiplexer and a two-mode homodyne detector whose classical Fisher information (CFI) approaches the QFI of the squeezed-state probe as the total probe state energy increases. We close the paper with a conjecture that the optimum Gaussian-state probe is a single-mode displaced squeezed state, discussions on the (non-Gaussian) QFI-achieving receiver, and extensions to multi-parameter estimation. Our results in this paper, together with our previous work on the optimum classical probe for beam displacement sensing~\cite{he2024optimum}, imply that in the far-field regime, the optimal single-mode Gaussian state probe approaches (classical) coherent state illumination. This observation, together with the results presented in this paper on the near-field quantum-optimum single-mode Gaussian-state probe, subsume the entire near-to-far field regimes of diffraction-limited propagation, for this problem.

\subsection{Outline of the paper}
The remainder of the paper is organized as follows. In section II, we introduce the free-space propagation channel enclosed by soft Gaussian apertures, and the singular value decomposition based on HG modes, which underlies the construction of the transverse displacement encoding unitary. Then we introduce a transformed modal basis to simplify the parameter-encoding unitary. In section III, we consider a single-spatial-mode squeezed-state probe in the deep-near-field regime (no loss), and show that the choice of the spatial mode determines whether the sensitivity is governed by the standard quantum limit (SQL) or the Heisenberg limit (HL). In section IV, we consider jointly optimizing the spatial mode to excite a squeezed state in, and the distribution of the squeezing and coherent energy proportions thereof, for a general return-path Fresnel number product, i.e., in the presence of diffraction loss. We show that in the limit of small beam displacement, the resulting infinite-spatial-mode, lossy, linear transformation can be represented compactly in a three-mode quantum circuit. This enables an efficient numerical solution to the aforesaid joint optimization problem. We show that this optimized single-spatial-mode squeezed-state probe achieves superior performance to the optimal-spatial-mode laser-light probe~\cite{he2024optimum} and that a two-spatial-mode homodyne receiver approaches the QFI of this quantum transmitter in the limit of high transmit probe energy. Section V concludes the paper with a discussion on future extensions and applications.
\begin{figure}
    \centering
\includegraphics[width=\linewidth]{ 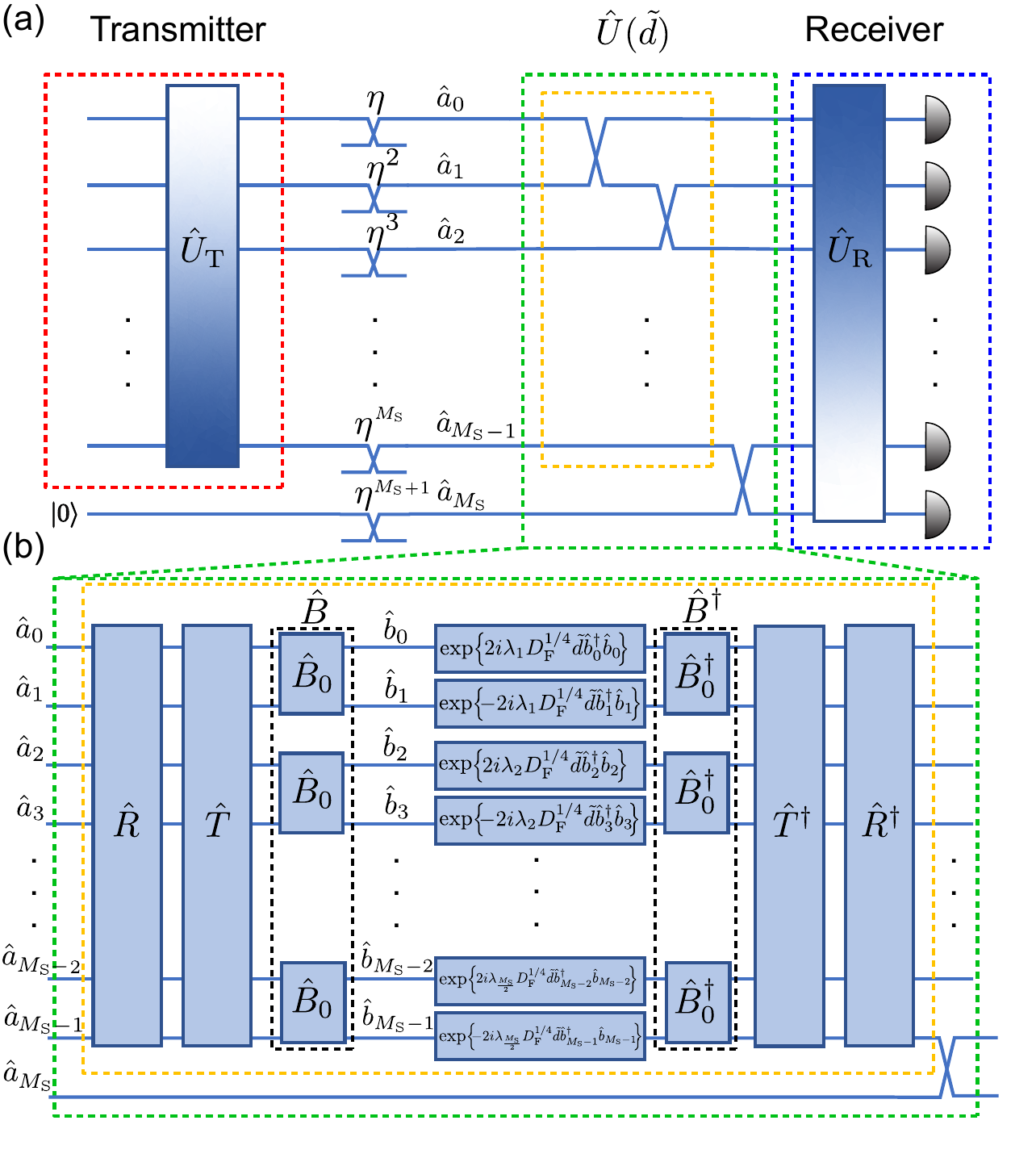}
\vspace{-5pt}
    \caption{(a) Circuit representation of a transmitter-receiver architecture in the HG basis. The parameter-encoding unitary transformation, $\hat{U}(\tilde{d})$ (in the dashed green box), is composed of consecutive beam splitters for all the neighboring HG modes. The field operator for the HG$_{n,0}$ mode at the receiver is $\hat{a}_n$. The passive unitary transformations at the transmitter and receiver, $\hat{U}_\mathrm{T(R)}$ (in gradient shade), represent reconfigurable spatial mode transformations. (b) After parameter-independent modal basis rotation, $\hat{U}(\tilde{d})$ is represented as independent single-mode phases in mode $\vec{\hat{b}}$, with the phase-evolution strength decreasing as the mode order increases.}
    \label{fig:2}
    \vspace{-10pt}
\end{figure}
\section{Problem setup}

\subsection{Propagation Geometry}
Fig.~\ref{fig:1}(a) presents a schematic of {\em optical beam deflectometry}, in which the angular displacement of a surface is encoded in the transverse displacement $d$ of a reflected probe beam. Assuming the surface is large compared to the spot size of the beam, one can unfold the beam propagation path into a line-of-sight free-space propagation channel depicted in Fig.~\ref{fig:1}(b), where $L$ is the return-path propagation distance and $r_\mathrm{T}$ and $r_\mathrm{R}$ are the radii of the transmitter- and receiver-side apertures of the imaging system, respectively. We assume Gaussian soft-aperture pupils at both ends with transverse amplitude-transmittance profiles $\mathcal{A}_{\mathrm{T(R)}}(x,y)=e^{-(x^2 + y^2)/r_\mathrm{T(R)}^2}$. A singular value decomposition of the propagation kernel---including the soft-aperture attenuation at both ends with Fresnel diffraction over $L$ meters in between---yields the infinite Laguerre Gauss (LG) and the Hermite Gauss (HG) spatial mode sets as the bi-fold degenerate {\em normal modes}~\cite{he2021QKDperformance,shapiro2005ultimate}. In other words, if only the HG mode $\Phi_{n}(x) \Phi_{l}(y)$ at the transmitter side is excited ($l,n \in \left\{0, 1, \ldots, \infty\right\}$), and there is no beam displacement ($d=0$), then only the receiver-side HG mode $\phi_{n}(x)\phi_{l}(y)$ is excited, up to a power-transmissivity $\eta^{n+l+1}$, where $\eta=(1+2D_\mathrm{F}-\sqrt{1+4 D_\mathrm{F}})/(2D_\mathrm{F})$ is the transmissivity of the fundamental (Gaussian) mode. In this paper, we will be primarily concerned with the {\em near-field propagation regime}, which is characterized by the Fresnel number product $D_\mathrm{F} \equiv (k r_\mathrm{T} r_\mathrm{R} /4L)^2 \gg 1$, where $k$ is the wavenumber ($2\pi$ divided by the center wavelength of the probe light). In the near-field regime the channel can accommodate roughly $D_{\rm F}$ mutually-orthogonal spatial modes with near-unity transmissivity. In the event of a beam displacement $d > 0$, the transmitter- and receiver-side HG modes cease to be independent parallel single-mode channels. 

\subsection{Crosstalk induced by beam displacement}
We define a {\em crosstalk matrix} $Q({\tilde d})$, where $\tilde{d} \equiv d/r_\mathrm{R}$ is the beam displacement normalized to the receiver-aperture radius, whose $(m, n)$-th element equals the amplitude-crosstalk between the receiver-side modes $\phi_{m}(x) \phi_{0}(y)$ and $\phi_{n}(x+d) \phi_{0}(y)$, namely,
\begin{equation}
\begin{aligned}
Q(\tilde{d})_{m,n}&\equiv\iint\displaylimits_{-\infty}^{\infty}\phi_{m}^{\ast}\left(x\right)\phi_{0}^{\ast}\left(y\right)\phi_{n}(x+d) \phi_{0}(y) \\
&\times\frac{\mathcal{A}_\mathrm{R}(x,y)}{\mathcal{A}_\mathrm{R}(x+d,y)} \mathrm{d}x\mathrm{d}y.
\end{aligned}
\end{equation}
$Q(\tilde{d})$ represents a passive unitary transformation over an infinite spatial-mode basis at the receiver, induced by a small beam displacement $\tilde d$. Performing the integral in the ${\tilde d} \ll 1$ limit~\cite{he2024optimum}, we obtain:
\begin{equation}
\begin{aligned}
Q(\tilde{d})
\approx&\;\boldsymbol{\mathcal{I}}+\tilde{d}\left[(1+4D_\text{F})^{1/4} \Gamma+\frac{ |\Gamma|}{(1+4D_\text{F})^{1/4}}\right.\\
&\left.+\frac{i r_\text{R} 2 \sqrt{D_\text{F}}}{r_\text{T} (1+4D_\text{F})^{1/4}} |\Gamma|\right],\\
\approx &\;\boldsymbol{\mathcal{I}}+\tilde{d}\left[(1+4D_\text{F})^{1/4} \Gamma+\frac{i r_\text{R} 2 \sqrt{D_\text{F}}}{r_\text{T} (1+4D_\text{F})^{1/4}} |\Gamma|\right],\\
\approx&\;\boldsymbol{\mathcal{I}}+(4 D_\mathrm{F})^{1/4}\tilde{d} \left[\Gamma+i|\Gamma|\right],
\end{aligned}
\label{eq:crosstalkmatrix_approx}
\end{equation}
where $\boldsymbol{\mathcal{I}}$ is an identity matrix and $\Gamma$ is a tridiagonal skew symmetric matrix~\cite{he2024optimum}, whose $M_\mathrm{S}\times M_\mathrm{S}$ dimensional leading principal submatrix is given by:
\begin{equation}
\Gamma\left(M_{\mathrm{S}}\right)\equiv\left[\begin{array}{ccccc}
0 & 1 & 0 & \ldots & 0 \\
-1 & 0 & \sqrt{2} & \ldots & 0 \\
0 & -\sqrt{2} & 0 & \ddots & \vdots \\
\vdots & \vdots & \ddots & \ddots & \sqrt{M_{\mathrm{S}}-1} \\
0 & 0 & \ldots & -\sqrt{M_{\mathrm{S}}-1} & 0
\end{array}\right].
\end{equation}
The approximation in the second row of Eq.~\eqref{eq:crosstalkmatrix_approx} holds in the $D_\mathrm{F} \gg 1$ limit. In the first row of Eq.~\eqref{eq:crosstalkmatrix_approx}, the third term is smaller than the second and fourth terms by a factor $(1+4 D_\text{F})^{1/2}\approx 2 \sqrt{D_\text{F}}$. When $D_\mathrm{F}=100$ this ratio is more than $20$. The approximation in the third row of Eq.~\eqref{eq:crosstalkmatrix_approx} holds in the $D_\mathrm{F} \gg 1$ limit and assuming $r_\mathrm{T}=r_\mathrm{R}=r_0$. 

\subsection{The Hamiltonian description}
We assume the transmitter has an even modal support $M_{\mathrm{S}}$, meaning we are allowed to excite the linear-span of the modes HG$_{0,0}$ through $\mathrm{HG}_{M_\mathrm{S}-1,0}$ in any quantum state, and $M_{\mathrm{S}}$ is an even integer. We denote the relevant crosstalk matrix under this transmitter constraint, i.e., for the first $M_\mathrm{S}+1$ modes, with the $M_\mathrm{S}+1$-th transmitter mode in a vacuum state, as:
\begin{equation}
    Q_1(\tilde{d})\equiv \boldsymbol{\mathcal{I}}(M_\mathrm{S}+1)+(4 D_\mathrm{F})^{\frac{1}{4}}\tilde{d} \left[\Gamma(M_\mathrm{S}+1)+i|\Gamma(M_\mathrm{S}+1)|\right],
\end{equation}
which is a skew Hermitian matrix with dimensions $(M_\mathrm{S}+1)\times (M_\mathrm{S}+1)$. The corresponding unitary transformation is $\hat{U}(\tilde{d})=\exp{-{\vec{\hat{a}}}^\dagger \ln{\left(Q_1({\tilde d})\right)}\vec{\hat{a}}} \equiv \exp{-i \tilde{d} \hat{H}}$, where $\vec{\hat{a}}$ is a column vector comprising of annihilation operators $\hat{a}_n$, $0 \le n \le M_\mathrm{S}$, associated with the receiver-side $\phi_n(x)\phi_0(y)$ modes. The dagger on the vector ${\vec{\hat{a}}}$ denotes the transpose of the vector and the Hermitian conjugate applied element-wise. The Hamiltonian $\hat{H}$ decomposes into pairwise beam-splitter interactions between neighboring HG modes with evolution strength increasing with mode order, 
\begin{equation}
\begin{aligned}
        \hat{H} \approx &-i \vec{\hat{a}}^\dagger \frac{1}{\tilde{d}}\left(Q_1(\tilde{d})-\boldsymbol{\mathcal{I}}(M_\mathrm{S}+1)\right)\vec{\hat{a}},\\
    =& 2 D_\mathrm{F}^{\frac{1}{4}} \sum\displaylimits_{n=0}^{M_\mathrm{S}-1} \sqrt{n+1} \left(e^{-\frac{i\pi}{4}}\hat{a}_{n+1}\hat{a}_{n}^\dagger+e^{\frac{i\pi}{4}}\hat{a}_{n}\hat{a}_{n+1}^\dagger\right),
\end{aligned}
\label{eq:Ha_main}
\end{equation}
where the approximation in the first equation above results from assuming ${\tilde d} \ll 1$. 

\subsection{Bosonic circuit decomposition}
A general quantum circuit representation of the transceiver model is shown in Fig.~\ref{fig:2}(a). $\hat{U}(\tilde{d})$, enclosed in the green box, is the receiver-side unitary that encodes the small beam displacement $\tilde{d}$. The yellow-boxed portion of $\hat{U}(\tilde{d})$ is what acts on the first $M_\mathrm{S}$ modes (in which the probe is excited). With this beam displacement accounted for (undone), the propagation of the individual HG modes---normal modes of the free-space propagation kernel without beam displacement---as independent single-mode pure-loss channels of transmissivities $\eta^{n+1}$, $0 \le n \le M_\mathrm{S}$. $\hat{U}_{\mathrm{T}}$ represents a potentially reconfigurable spatial mode transformation enabling the excitation of the linear span of first $M_\mathrm{S}$ transmitter-side HG modes in a Gaussian entangled state by linear mixing of a collection of single-mode squeezed states. In a similar vein, we include $\hat{U}_{\mathrm{R}}$ on the receiver side as a potentially reconfigurable spatial mode transformation to enable measurement in the most information-rich modes. 

Fig.~\ref{fig:2}(b) shows an equivalent representation of $U(\tilde d)$ that eases the analysis of the QFI-optimal transmitter. The $\tilde d$ dependence in this new representation---a collection of $M_{\mathrm S}/2$ phase-conjugate interferometers flanked by multimode passive unitaries---appears only as independent single-mode phases of the form $\exp
[(-1)^{n+1}2i\lambda_{\lfloor{n/2}\rfloor +1} D_{\rm F}^{1/4}{\tilde d}{\hat b_n^\dagger}{\hat b_n}]$ acting on a set of modes $\vec{\hat{b}}\equiv\hat{B} \hat{T} \hat{R} \vec{\hat{a}}$, which are a linear transformation of the modes $\vec{\hat{a}}$ via parameter-independent unitaries $\hat{R}$, $\hat{T}$ and $\hat{B}$. The eigenvalues of $\Gamma(M_\mathrm{S})$ are imaginary and occur in pairs $\pm i\lambda_n$ for $1 \le n \le M_\mathrm{S}/2$, ordered decreasingly $\lambda_1 \ge \lambda_2 \ge \ldots \ge \lambda_{M_\mathrm{S}/2}$. The derivation of the representation in Fig.~\ref{fig:2}(b) in Appendix~\ref{app:parameter_encoding} shows that $\hat{B}_0$ is a $50:50$ beam splitter, $\hat{T}$ is an orthogonal transformation, and $\hat{R}$ consists only of single-mode phases. In terms of $\vec{\hat{b}}$, the Hamiltonian $\hat H$ is therefore given by:
\begin{equation}
\begin{aligned}
    \hat{H}=&\sum \displaylimits_{n=1}^{M_\mathrm{S}/2} -2  \lambda_n D_\mathrm{F}^{\frac{1}{4}} \left(\hat{b}_{2n-2}^\dagger\hat{b}_{2n-2}-\hat{b}_{2n-1}^\dagger\hat{b}_{2n-1}\right)\\
    &+2 D_\mathrm{F}^{\frac{1}{4}}\sum\displaylimits_{n=0}^{M_\mathrm{S}-1}\left(e^{-\frac{i \pi}{4}} \hat{a}_{M_\mathrm{S}}\hat{b}_{n}^\dagger e ^{-i\varrho_n}+e^{\frac{i \pi}{4}} e ^{i\varrho_n}\hat{b}_{n}\hat{a}_{M_\mathrm{S}}^\dagger\right).
\end{aligned}
\label{eq:Hb_main}
\end{equation}
where $\varrho_n$ is the complex phase of the $n$-th entry of the last ($M_\mathrm{S}$-th) row vector of $R^\ast T^T B^\dagger$. 

\section{Analysis in the lossless limit}
We examine the channel shown in Fig.~\ref{fig:2} in the lossless limit. In other words, we consider the regime $M_\mathrm{S} \ll D_{\rm F}$, such that the first $M_\mathrm{S}$ modes---in the linear span of which the probe light will be assumed to be contained---are effectively lossless. We consider a single-spatial-mode probe, and show that if the excited mode has a fixed parity (viz., it is a strictly even or an odd function), even a quantum (squeezed-state) excitation of that mode cannot attain Heisenberg Limited (HL) sensitivity. However, we will show that a squeezed state of a spatial mode without a fixed parity can attain HL sensitivity.

\subsection{An intuitive argument}

Let us first consider the following argument that will help guide our intuition for the results of this section. Ref.~\cite{matsubara2019optimal} proves that the QFI-optimal Gaussian-state probe to estimate a single parameter encoded in a multimode passive unitary is a single-mode squeezed vacuum in an appropriate mode in the modal span on which the unitary acts. A single-mode homodyne receiver was shown to achieve this optimum QFI. The yellow-boxed portion of Figs. \ref{fig:2}(a) and (b) is a unitary, and hence---with a transmitter restricted to employ the first $M_{\rm S}$ modes---the results of Ref.~\cite{matsubara2019optimal} may appear to hold here. However, since a small ${\tilde d} > 0$ spills over photons into the $(M_{\rm S} + 1)$-th mode at the receiver (the last output of the green box in Fig. \ref{fig:2}(b)), Ref.~\cite{matsubara2019optimal} does not strictly apply to our problem. If we were to pretend that we only have access to the first $M_{\rm S}$ receiver-side modes, the results of Ref.~\cite{matsubara2019optimal} applies. Under this (hypothetical) restriction, given the decreasing magnitudes of $\lambda_m$, $1 \le m \le M_\mathrm{S}/2$, and hence the diminishing phase strengths modulating the $\hat b$ modes, under a total mean photon number constraint, exciting the ${\hat b}_0$ (or the ${\hat b}_1$) mode in a single-mode squeezed-vacuum is QFI optimal. The ${\hat b}_0$ (or, the ${\hat b}_1$) mode, transformed back to the $\left\{\hat a\right\}$ modes reveals the transmitter-side mode(s) which when excited in the single-mode squeezed vacuum achieves the optimum QFI in this restricted setting. We will now evaluate the QFI when the transmitter excites the ${\hat b}_0$ (or the ${\hat b}_1$) mode in a squeezed vacuum state, but without the above said restriction, in the lossless limit ($M_\mathrm{S} \ll D_{\rm F}$). We will show that the above intuition from the restricted setting, and the associated HL scaling of sensitivity of estimating $\tilde d$, holds true, but in the limit of high probe energy. This is because the QFI contribution from the photons in the $(M_{\rm S}+1)$-th receiver mode bears an SQL scaling, and is masked by the HL scaling QFI contributed by the photons in the first $M_{\rm S}$ receiver-side modes.

\subsection{QFI attained by exciting a high-order HG mode in a squeezed state: SQL scaling sensitivity}
Let us consider a single-spatial-mode pure Gaussian state, i.e., the  {\em displaced-squeezed} state $\ket{\boldsymbol{\alpha};r e^{i\vartheta_1}}$, of the $\mathrm{HG}_{n,0}$ mode $\Phi_n(x)\Phi_0(y)$, tensor product with the vacuum state of every HG-basis mode other than ${\rm HG}_{n,0}$, where $\boldsymbol{\alpha} \in {\mathbb C}$ represents a phase-space displacement, $r \ge 0$ is the squeezing parameter and $0\leq\vartheta_1<\pi$ is the squeezing angle. The total mean photon number of this displaced-squeezed state is given by: 
\begin{equation}
|\boldsymbol{\alpha}|^2+\mathrm{sinh}^2(r) \equiv N.
\end{equation}
When $r=0$, this state reduces to a coherent state. When $\boldsymbol{\alpha}=0$, it reduces to a squeezed vacuum state. The model developed in the previous section implies that the modes relevant for analyzing the effect of a small transverse displacement are the following three receiver-side modes: $\mathrm{HG}_{n,0}$ and the two {\em adjacent} modes $\mathrm{HG}_{n \pm 1,0}$. Hence, we can truncate the multimode Hamiltonian in Eq.~\eqref{eq:Ha_main} to:
\begin{equation}
\begin{aligned}
    \hat{H}_1=&2 D_\mathrm{F}^{\frac{1}{4}} \left\{ \sqrt{n} \left(e^{-\frac{i\pi}{4}}\hat{a}_{n}^{}\hat{a}_{n-1}^\dagger+e^{\frac{i\pi}{4}}\hat{a}_{n-1}^{}\hat{a}_{n}^\dagger
    \right)\right.\\
    &\left.+\sqrt{n+1} \left(e^{-\frac{i\pi}{4}}\hat{a}_{n+1}^{}\hat{a}_{n}^\dagger+e^{\frac{i\pi}{4}}\hat{a}_{n}^{}\hat{a}_{n+1}^\dagger
    \right) \right\}.
\end{aligned}
\end{equation}  
This Hamiltonian only contains the two beam splitter interactions between $\mathrm{HG}_{n,0}$ and its adjacent modes. The QFI for a pure-state probe $|\omega\rangle$ under a Hamiltonian ${\hat H}$, $K[\ket{\omega},{\hat H}]$, is given by $4 \left<\Delta^2\hat{H} \right>_{|\omega\rangle}$ ~\cite{braunstein1994QFI}, which yields:
\begin{equation}
\label{eq:K1}\begin{aligned}
    K\left[|0\rangle \ket{\boldsymbol{\alpha};r e^{i\vartheta_1}}|0\rangle,\hat{H}_1\right] \equiv& 4 \left<\Delta^2\hat{H}_1 \right>_{|0\rangle \ket{\boldsymbol{\alpha};r e^{i\vartheta_1}}|0\rangle},\\
   =&16 \sqrt{D_\mathrm{F}} (2n+1) N,
\end{aligned}
\end{equation}
where the state ${|0\rangle \ket{\boldsymbol{\alpha};r e^{i\vartheta_1}} |0\rangle}$ is of the modes HG${_{n-1,0}}$, HG${_{n,0}}$ and HG${_{n+1,0}}$ respectively, with the detailed evaluation of the RHS being relegated to Appendix~\ref{app:K_calc}. This implies that the QFI attained by a single-mode Gaussian-state probe of mean photon number $N$ of any ${\rm{HG}}_{n,0}$ mode is independent of the squeezing parameter $r$, and thereby attained by a coherent state $|\boldsymbol{\alpha};0\rangle$, which in turn follows the standard quantum limited (SQL) scaling (QFI $\propto N$)~\cite{quantummetrology}. The QFI also is seen to scale linearly in the HG mode order $n$, indicating the superior sensitivity of higher-order modes, modulo the no-loss assumption stemming from the $M_{\rm S} \ll D_{\rm F}$ premise underlying this section. A receiver that saturates the QFI is single-mode homodyne detection in an appropriately weighted linear combination of the $\mathrm{HG}_{n-1,0}$ and $\mathrm{HG}_{n+1,0}$ modes~\cite{he2024optimum}.
\begin{figure}
    \centering
\includegraphics[width=1\linewidth]{ 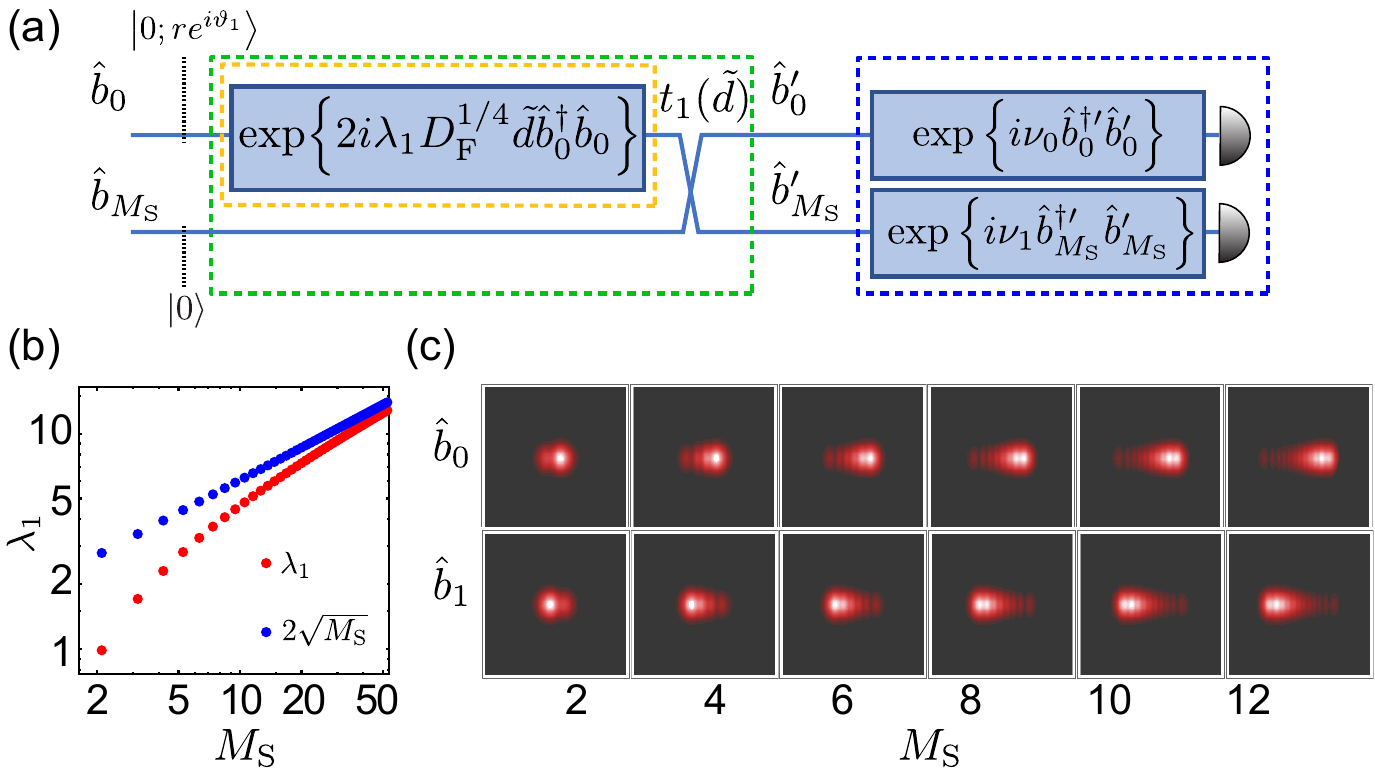}
    \caption{(a) parameter-encoding circuit in the lossless limit for a single-mode state excited in $\hat{b}_0$ is enclosed in a green box. The beam splitter has a transmissivity: 
$t_1(\tilde{d})=\cos^2{(2D_\mathrm{F}^{\frac{1}{4}}\tilde{d})}$. A two-mode homodyne measurement with $\nu_0=\mathrm{ArcTan(e^{2r})}$ and $\nu_1=\frac{\pi}{4}$ is enclosed in a blue box. (b) Largest phase evolution strength $\lambda_1$ versus the modal support at the transmitter, $M_\mathrm{S}$. (c) Intensity profile of modes $\hat{b}_0$ and $\hat{b}_1$ for different modal support. }
    \label{fig:3}
    \vspace{-10pt}
\end{figure}

\subsection{QFI attained by exciting the ${\hat b}_0$ mode in a squeezed-vacuum state: emergence of HL scaling}

Next, we evaluate the QFI of a squeezed vacuum state $\ket{0;r e^{i\vartheta_1}}$ of the mode $\hat{b}_{0}$, which is a linear superposition of all the transmitter-side $M_{\rm{S}}$ modes (see Fig.~\ref{fig:2}), and lacks a fixed parity. The first summation in the Hamiltonian in Eq.~\eqref{eq:Hb_main} represents single-mode phases accrued on modes ${\hat b}_{2n-2}$ and ${\hat b}_{2n-1}$ respectively, whereas the second summation shows that if the probe only excites the ${\hat b}_0$ mode, we need only consider the term in that summation corresponding to interactions between the two modes ${\hat b}_0$ and ${\hat a}_{M_\mathrm{S}}$. Therefore, without loss of generality, we can reduce the Hamiltonian in Eq.~\eqref{eq:Hb_main} to:
\begin{equation}
\begin{aligned}
\label{eq:H_bmode}
    \hat{H}_2=&-2  \lambda_1 D_\mathrm{F}^{\frac{1}{4}} \hat{b}_{0}^\dagger\hat{b}_{0}^{}\\
&+2 D_\mathrm{F}^{\frac{1}{4}} \left(e^{-\frac{i \pi}{4}} \hat{a}_{M_\mathrm{S}}^{}\hat{b}_{0}^\dagger e ^{-i\varrho_0}+e^{\frac{i \pi}{4}} e ^{i\varrho_0}\hat{b}_{0}^{}\hat{a}_{M_\mathrm{S}}^\dagger\right),\\
    =&-2  \lambda_1 D_\mathrm{F}^{\frac{1}{4}} \hat{b}_{0}^\dagger\hat{b}_{0}^{}+2 D_\mathrm{F}^{\frac{1}{4}}\left(\hat{b}_{M_\mathrm{S}}^{}\hat{b}_{0}^\dagger+\hat{b}_{0}^{}\hat{b}_{M_\mathrm{S}}^\dagger\right).
\end{aligned}
\end{equation}
In writing the second equality above, we leveraged the fact that the $\hat{a}_{M_\mathrm{S}}$ mode is in the vacuum state, to replace $e^{-\frac{i \pi}{4}} e ^{-i\varrho_0}\hat{a}_{M_\mathrm{S}}$ with a field operator $\hat{b}_{M_\mathrm{S}}$. This implies that the effect of a small transverse displacement reduces to a single-mode phase evolution of the ${\hat b}_0$ mode followed by a beam splitter interaction with a vacuum state of the mode $\hat{b}_{M_\mathrm{S}}$, as illustrated by the two-mode quantum circuit in Fig.~\ref{fig:3}(a). We show in Appendix~\ref{app:K_calc} that the QFI attained by the squeezed-vacuum state $|0; r e^{i\vartheta_1}\rangle$ of the ${\hat b}_0$ mode, $ {\rm{sinh}^2(r)}=N$, (and $\hat{b}_{M_\mathrm{S}}$ in the vacuum state) is given by:
\begin{equation}\label{eq:b1qfi}
\begin{aligned}
 K\left[\ket{0;r e^{i\vartheta_1}}\ket{0},\hat{H}_2\right]\equiv &4 \left<\Delta^2\hat{H}_2 \right>_{\ket{0;r e^{i\vartheta_1}}\ket{0}},\\
 =&32 D_\mathrm{F}^{\frac{1}{2}} \lambda_1^2 N(1+N)+ 16 D_\mathrm{F}^{\frac{1}{2}} N,
\end{aligned}
\end{equation}
which clearly exhibits HL scaling (QFI $\propto N^2$) in the large-$N$ limit. The origin of the first term (HL) in Eq.~\eqref{eq:b1qfi} is the excited ${\hat b}_0$ mode undergoing single-mode phase evolution (see first term of Eq.~\eqref{eq:H_bmode}), whereas the origin of the SQL scaling in the second term of Eq.~\eqref{eq:b1qfi} lies in the beam-splitter interaction of the probe mode ${\hat b}_0$ with the vacuum state of the ${\hat b}_{M_\mathrm{S}}$ mode in the second term of Eq.~\eqref{eq:H_bmode}. Furthermore, the increased sensitivity of higher-order spatial modes manifests itself in Eq.~\eqref{eq:b1qfi} through $\lambda_1$, which asymptotically scales as $\sim \sqrt{M_\mathrm{S}}$~\cite{qi2018ultimate}, as shown in Fig.~\ref{fig:3}(b). In Appendix~\ref{app:Lossless_twomode_homodyne} we prove that a two-mode homodyne detection receiver saturates the QFI, as pictorially depicted in Fig.~\ref{fig:3}(a).

The circuit in Fig.~\ref{fig:2} indicates that a single-mode probe that excites the $\hat{b}_1$ mode in a squeezed vacuum state has the same QFI as in Eq.~\eqref{eq:b1qfi}. In Fig.~\ref{fig:3}(c), we plot the spatial intensity profiles of modes $\hat{b}_0$ and $\hat{b}_1$, respectively, as a function of $M_\mathrm{S}$. The center of mass of this comet-like intensity pattern moves away from the origin as $M_\mathrm{S}$ increases. A single-mode squeezed-vacuum state of the ${\hat b}_0$ (or ${\hat b}_1$) mode, represented in the HG basis, is an $M_\mathrm{S}$-mode Gaussian entangled state~\cite{weedbrook2012gaussian}.  
The analysis of the above two cases suggests that the QFI attainable by a single-mode state for the estimation of a small transverse displacement is directly linked to the parity of the spatial mode in which the probe state is excited. For a spatial mode with a fixed parity (even or odd), the relevant evolution is a beam splitter interaction between the excited mode and vacuum, leading to a QFI that follows SQL scaling. For a spatial mode without a fixed parity, the relevant evolution is a single-mode phase applied to the excited mode and a beam splitter interaction between the excited mode and vacuum, where the phase evolution leads to a QFI that follows the HL scaling. Appendix~\ref{app:optmodecn} provides further physical intuition on the QFI-optimizing spatial mode in the lossless limit considered in this section.

Finally, in this near-lossless setting ($M_S \ll D_{\rm F}$), a numerical analysis of the circuit shown in Fig.~\ref{fig:2}(b) suggests that in the large-$N$ limit, a single-mode squeezed vacuum state of the $\hat{b}_{0}$ (or $\hat{b}_{1}$) mode is not only a HL-scaling-achieving probe as proved above, but is the QFI-optimal single-mode Gaussian-state probe. This empirical finding is in agreement with the HL scaling in the large-$N$ limit of the QFI expression shown in Eq.~\eqref{eq:b1qfi}.

\section{Optimal single-mode Squeezed-state probe: Analysis including diffraction loss}
\begin{figure}
    \centering 
    \includegraphics[width=1.\linewidth]{ 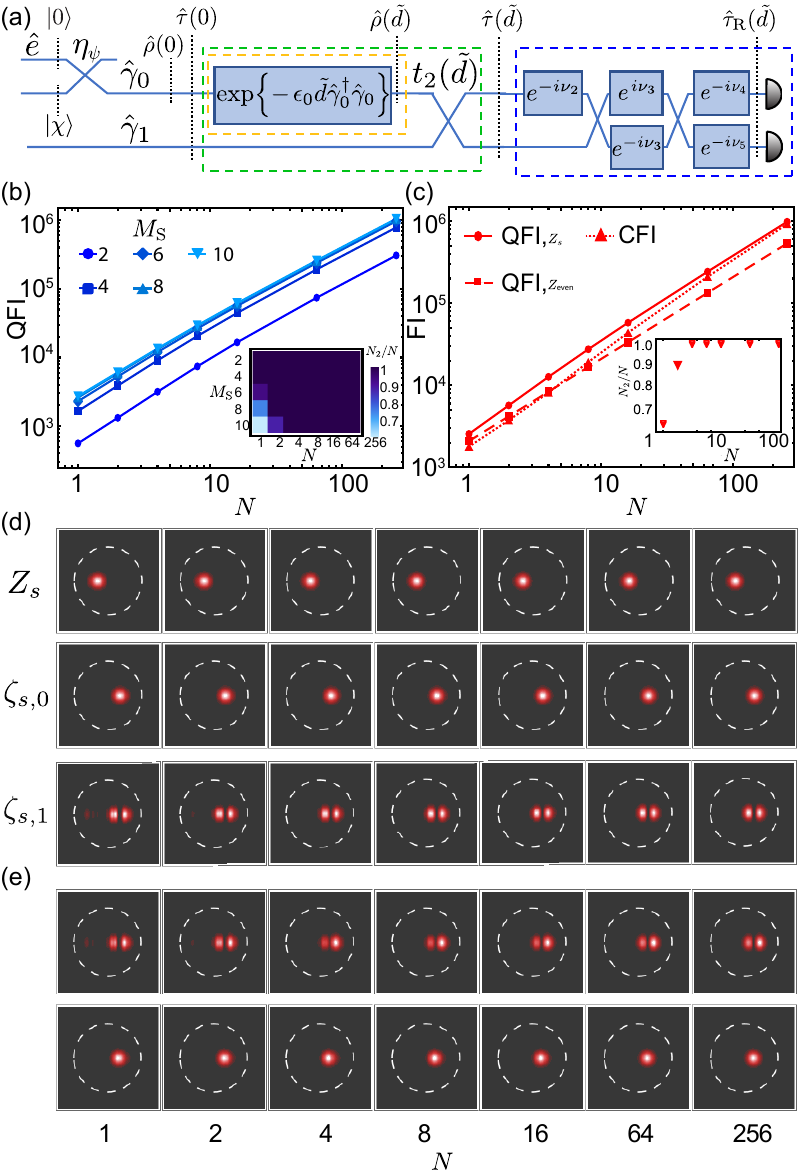}
    \caption{(a) Circuit model for transverse displacement estimation with a single-mode transmitter and a two-mode homodyne receiver. The parameter-encoding beam splitter has a transmissivity: $t_2(\tilde{d})=\cos^2(\sqrt{\epsilon^2-|\epsilon_0|^2 }\tilde{d})$. (b) The optimized QFI for a system with $D_\mathrm{F}=100$ versus mean photon energy $N$, for different transmitter modal support $M_\mathrm{S}$. (c) The optimized QFI for $M_\mathrm{S}=10$ (joint) from (b), the corresponding optimized CFI for a two-mode homodyne receiver (dotted), and the optimal coherent state's QFI (dashed), versus $N$. Insets show the fraction of the total photon energy attributed to the squeezing portion. (d) Intensity distributions of the optimal transmitter mode ($Z_s$) and the corresponding receiver modes of the principal component basis $\left\{\zeta_{s,0}, \zeta_{s,1}\right\}$ for $M_\mathrm{S} = 10$ and various $N$. (e) Intensity distributions of the optimal spatial modes to perform the two-mode homodyne in for the optimal single-mode Gaussian-state probe excited in $Z_s$ from panel (b).
     }
\label{fig:4}
    \vspace{-10pt}
\end{figure}
\subsection{The three-mode interaction picture}
 In the presence of loss, quantum resources no longer provide HL scaling of squared-error precision, but only a constant factor of enhancement over that of a classical probe~\cite{elusive-H-limit}. In this section, we will analyze a beam displacement system with Gaussian-aperture geometries and line-of-sight propagation distance, as shown in Fig.~\ref{fig:1}, while fully accounting for spatial-mode-dependent diffraction losses in the free-space propagation. We will limit our attention to finding the optimal single-spatial-mode Gaussian (displaced-squeezed) probe state. For this purpose, the general parameter-encoding circuit shown in Fig.~\ref{fig:2} is cumbersome to deal with. Starting with an arbitrary spatial mode $\Psi_0(x,y)$ that the probe excites in the quantum state $|\chi\rangle$, the free-space propagation and the beam-displacement parameter-encoding process can be encapsulated by a simple three-mode quantum circuit shown in Fig~\ref{fig:4}(a), where $\eta_\psi$ denotes the transmissivity of the mode $\Psi_0$, and the green-dashed box denotes a two-mode parameter-encoding circuit that is a function of $\tilde d$ and two evolution-strength parameters $(\epsilon,\epsilon_0)$, defined later in this section, where the beam splitter transmissivity, $t_2(\tilde{d})=\cos^2(\sqrt{\epsilon^2-|\epsilon_0|^2 }\tilde{d})$. This circuit leads to a straightforward calculation of the QFI as a function of the single spatial mode the probe excites and the quantum state of that mode. Using QFI as the merit function, we find the optimal single-mode squeezed-state probe, i.e., the energy allocation between the phase-space displacement and squeezing, and the QFI-optimal spatial mode to excite in the displaced-squeezed state. 

We excite an arbitrary transmitter side spatial mode, $\Psi_0(x,y)$ in a pure Gaussian quantum state $\ket{\chi}$, with
\begin{equation}
\Psi_0(x,y)=\sum_{n=0}^{M_\mathrm{S}-1} c_n e^{i \theta_n} \Phi_{n}(x) \Phi_{0}(y),
\end{equation}
where the {\em mode coefficients} $c_n, \theta_n \in \mathbb{R}$ and modal normalization is ensured by setting $\sum\displaylimits_{n=0}^{M_\mathrm{S}-1} c_n^2=1$. The diffraction-limited transmissivity of mode $\Psi_0$ is given by $
\eta_\psi=\sum\displaylimits_{n=0}^{M_\mathrm{S}-1}c_n^2 \eta^{n+1}$, which is a weighted sum of the transmissivities of the first $M_\mathrm{S}$ $\rm{HG}$ modes. This diffraction loss can be captured by a beam splitter interaction with an environment mode $\hat{e}$ in vacuum, as shown in Fig.~\ref{fig:4}(a). At the receiver, when there is no displacement (${\tilde d} = 0$), all the photons are in the mode:
\begin{equation}
    \psi_0(x,y)=\frac{1}{\sqrt{\eta_\psi}}\sum_{n=0}^{M_\mathrm{S}-1} c_n e^{i \theta_n} \eta^{(n+1)/2}\phi_{n}(x)\phi_{0}(y).
\end{equation} 
The excited receiver-side spatial mode in the presence of beam displacement $\tilde{d}\ll1$ is $ \frac{\psi_0(x-d,y)}{\mathcal{A}_{\mathrm{R}}(x-d,y)} \mathcal{A}_{\mathrm{R}}(x,y)$, where $\mathcal{A}_{\mathrm{R}}(x,y)$ is the amplitude-transmittance of the receiver pupil. Expanding it in Taylor series and using $\tilde{d}=d/r_\mathrm{R}$, 
\begin{equation}\label{eq:linearexpansion}
 \frac{\psi_0(x-d,y)}{\mathcal{A}_{\mathrm{R}}(x-d,y)} \mathcal{A}_{\mathrm{R}}(x,y) = \psi_0(x,y)+\epsilon \tilde{d} \psi_1^{\mathrm{aux}}(x,y)+\mathcal{O}(\tilde{d}^2),
\end{equation}
where $\epsilon$ ensures $\psi_1^{\mathrm{aux}}$ is a properly-normalized mode function, and it represents the evolution strength of the transformation induced by $\tilde{d}$. Applying the Gram–Schmidt process to the span of $\{\psi_0, \psi_1^{\mathrm{aux}}\}$ yields a mode $\psi_1$ that is orthogonal to $\psi_0$. We term this two-mode set, $\{\psi_0,\psi_1\}$, as the \textit{principal component} (PC) basis. We further define $\epsilon_0$ as the mode-overlap integral projecting $\epsilon \psi_1^{\mathrm{aux}}$ onto $\psi_0$, i.e., $\epsilon_1 \psi_1
\equiv \epsilon \psi_1^\mathrm{aux}(x,y) - \epsilon_0 \psi_0(x,y)$, with $\epsilon_1^2+|\epsilon_0|^2=\epsilon^2$.

In the PC basis, the linear expansion in Eq.\eqref{eq:linearexpansion} leads to a parameter-independent Hamiltonian, 
\begin{equation}
    \hat{H}_3=-i \epsilon_0 \hat{\gamma}_0^\dagger \hat{\gamma}_0+i \sqrt{\epsilon^2-|\epsilon_0|^2} (\hat{\gamma}_0^\dagger \hat{\gamma}_1-\hat{\gamma}_1^\dagger \hat{\gamma}_0),
    \label{eq:Hamiltonian_PCA}
\end{equation}
where $\{\hat{\gamma}_0, \hat{\gamma}_1\}$ represent the field operators of the receiver side modes $\{\psi_0,\psi_1\}$ respectively. See Appendix~\ref{app:lossy_channel} for further details. The corresponding unitary for $\hat{H}_3$ is $\hat{U}_3(\tilde{d})\equiv\exp{-i \tilde{d} \hat{H}_3}$, denoted by the green dashed box in Fig.~\ref{fig:4}(a). The first term of $\hat{H}_3$ captures the phase evolution of the mode $\psi_0$, and the second term describes the interaction between $\psi_0$ and the mode $\psi_1$ in vacuum. 

\subsection{QFI Optimization}
When the transmitter excites mode $\Psi_0$ in a single-mode pure Gaussian state $\ket{\chi}$, the receiver-side mode $\psi_0$ appears in a mixed state $\hat{\rho}(0)$. We use $\hat{\tau}(0)=\hat{\rho}(0)\otimes\ket{0}\bra{0}$ to represent the two-mode state of the receiver-side modes $\psi_0$ and $\psi_1$. Thus the parameter-encoded state, $\hat{\tau}(\tilde{d})=\hat{U}_3(\tilde{d})\hat{\tau}(0) \hat{U}_3^\dagger(\tilde{d})$. When a multimode state probes a passive linear-optical unitary that encodes a single parameter, the QFI for estimating the parameter can be expressed as the sum of the contribution from interactions among the modes that the probe excites, and the contribution from interactions between the excited modes and ancillary vacuum modes~\cite{2023mode_parameter_estimation}. In particular, for a single-mode state probing a multimode passive unitary, the parameter-encoding transformation is representable as a single-mode phase on the excited mode and a beamsplitter interaction between the excited mode and one ancilla mode in vacuum. Thus we have ~\cite{2023mode_parameter_estimation}:
\begin{equation}
\label{eq:qfi_mode_and_state}
K\left[\hat{\tau}(0),\hat{H}_3\right]=
K\left[\hat{\rho}(0), -i\epsilon_0\hat{N}_0\right]+4\left[\epsilon^2-|\epsilon_0|^2\right]\langle\hat{N}_0\rangle_{\hat{\rho}(0)},
\end{equation} 
where $\hat{N}_0=\hat{\gamma}_0^\dagger \hat{\gamma}_0$. The first term in the RHS of Eq.~\eqref{eq:qfi_mode_and_state} represents the QFI content associated with the phase evolution accrued on the state $\hat{\rho}(0)$. The second term in Eq.~\eqref{eq:qfi_mode_and_state} represents the QFI content from the beam splitter interaction between the excited mode $\psi_0$ and the mode $\psi_1$ in vacuum. This latter term is proportional to the total mean photon number in $\hat{\rho}$ where $\langle\hat{N}_0\rangle_{\hat{\rho}(0)}=\eta_\psi N$. 

 We evaluate $K\left[\hat{\rho}(0),-i\epsilon_0\hat{N}_0\right]$ using the mean vector and covariance matrix representation of a Gaussian state and the symplectic matrix representation of the Gaussian unitary transformations~\cite{weedbrook2012gaussian,pinel2013quantumsinglemodegaussian_single_parameter} involved in the three-mode circuit shown in Fig.~\ref{fig:4}. We summarize the results below. We start with the most general single-mode pure Gaussian state $\ket{\chi}=\ket{\boldsymbol{\alpha};r e^{i \vartheta_1}}$, where $\boldsymbol{\alpha}=\alpha e^{i\vartheta_0}$, $\alpha\in\mathbb{R}$, $0\leq\vartheta_0< 2\pi$, $0\leq\vartheta_1< \pi$ and $r\ge0$. We define two parameters: $r^\prime\equiv \frac{1}{4} \mathrm{ln}\left( \left(1-\eta_\psi+ e^{2r} \eta_\psi \right)\right.$ $/\left.\left( 1-\eta_\psi+ e^{-2r} \eta_\psi\right)\right)$ denoting the effective squeezing, and $v\equiv\sqrt{( 1-\eta_\psi+ e^{-2r} \eta_\psi)( 1-\eta_\psi+ e^{2r} \eta_\psi)}$ denoting the purity, to put the covariance matrix of $\hat{\rho}(0)$ into the thermal decomposition form~\cite{weedbrook2012gaussian}, which leads to~\cite{pinel2013quantumsinglemodegaussian_single_parameter} (see Appendix~\ref{app:lossyQFI} for detailed steps of the derivation):
\begin{equation}
\begin{aligned}\label{eq:QFI_singlemode_phase_main}
 K\left[\hat{\rho}(0), -i\epsilon_0\hat{N}_0\right]&=|\epsilon_0|^2\left(\frac{4 \mathrm{sinh}^2(2r^\prime)}{1+v^{-2}}+\frac{4}{v} \eta_\psi \alpha^2 e^{2r^\prime}\right).\\
\end{aligned}
\end{equation}
Thus far, $K\left[\hat{\tau}(0),\hat{H}_3\right]$ is represented as a function of the transmitted squeezed state $\ket{\chi}$'s phase-space displacement $\alpha$, squeezing $r$, and the spatial mode coefficients $\left\{c_n,\theta_n\right\}$. Appendix~\ref{app:simplification of merit function} presents explicit simplified formulas for the mode-dependent parameters $\left\{\epsilon_0, \epsilon,\eta_\psi\right\}$ purely in terms of $\left\{c_n\right\}$ and the geometrical parameters $D_{\rm F}$ and $r_{\rm R}$. Maximizing the QFI~\eqref{eq:qfi_mode_and_state} gives us the optimal single-mode Gaussian state probe parameterized solely by $\{M_\mathrm{S}, D_\mathrm{F}, N\}$. Namely, when $M_\mathrm{S}$ lowest-order HG modes are allowed to be excited at the transmitter for a system whose diffraction loss is quantified by $D_\mathrm{F}$, for a single-mode Gaussian state probe with total mean photon number $\alpha^2+\sinh^2(r)=N$, we find the mode coefficients $\{c_n,\theta_n\}$ of the optimal mode, and the optimal energy allocation between the dispalcement and squeezing components, viz., $\{N_1,N_2\}$ where $N_1\equiv\alpha^2$ and $N_2\equiv\sinh^2(r)$, for the QFI-optimal single-mode Gaussian state. The initial squeezing angle $\vartheta_0$ and the phase-space displacement angle $\vartheta_1$ do not change the QFI.

\subsection{Numerical Results and Discussion}
The optimized QFI for $D_\mathrm{F}=100$ is plotted as a function of $N$ in Fig.~\ref{fig:4}(b), with the inset showing optimal fractional energy allocation into the squeezing portion, $N_2/N$. The QFI increase saturates as $M_\mathrm{S}$ increases. With sufficient modal support for this optimum-QFI saturation, we denote the transmitter side \textit{optimal spatial mode} as $Z_s$ and the \textit{optimal single-mode Gaussian-state probe} as $\ket{\chi}_{\mathrm{opt}}$. For various $N$ values, the intensity distributions of $Z_s$, and the corresponding receiver modes in the PC basis, $\{\zeta_{s,0},\zeta_{s,1}\}$, are shown in Fig.~\ref{fig:4}(d). The optimal modes are highly asymmetrical, ensuring $|\epsilon_0| \neq 0$. Fig.~\ref{fig:4}(c) compares this QFI with the optimal QFI for a classical laser-light probe, and establishes that---unlike previous works that showed multi-spatial mode probes in the HG basis that provided quantum advantage---a {\em single-mode} squeezed-state probe, which is easier to prepare experimentally~\cite{squeezed_light_pump}, enables squeezing-enhanced sensitivity. Furthermore, no multimode Gaussian-state probe known to us exceeds the QFI performance of our optimized single-mode probe, for this problem. 

For $\ket{\chi}_{\mathrm{opt}}$, we optimize the CFI of a two-mode homodyne receiver. A general two-mode homodyne receiver can be parameterized with four variables, $\{\nu_2,\nu_3,\nu_4,\nu_5\}$, with the circuit representation shown in Fig.~\ref{fig:4}(a), enclosed by a blue dashed box, and the state right before detection denoted by $\hat{\tau}_\mathrm{R}(\tilde{d})$. We send the two-mode parameter-carrying state through a general two-mode Mach-Zehnder Interferometer (MZI)~\cite{clements2016optimal}, parameterized by $\{\nu_2,\nu_3\}$, to find the \textit{optimal measurement modes}. A pair of single-mode phases are then applied to the two modes separately, such that when the real quadratures of $\hat{\tau}_\mathrm{R}(\tilde{d})$ are measured, homodyne measurements along the chosen quadratures are performed. This measurement produces two random variables following multi-variate normal distribution with a mean vector $\overline{X}_1$ and a covariance matrix $\Sigma_1$~\cite{weedbrook2012gaussian}. The probability distribution is
\begin{equation}
    P_1(\mathbbm{q}_1,\mathbbm{q}_2)= \frac{\exp{-\frac{1}{2} (\boldsymbol{q}-\overline{X}_1) \Sigma_1^{-1}(\boldsymbol{q}-\overline{X}_1)^T}}{2\pi \mathrm{det}(\Sigma_1)^{1/2}},
\end{equation}
where $\boldsymbol{q}\equiv\{\mathbbm{q}_1,\mathbbm{q}_2\}$ is a vector. The CFI of a measurement is the expected value of the square of the log-likelihood function's first order derivative with respect to the parameter of interest, $E\left[(\partial(\ln P_1)/\partial \tilde{d})^2\right]=E\left[\left(\frac{\partial P_1}{\partial \tilde{d}}\frac{1}{P_1}\right)^2\right]$. For $D_\mathrm{F}=100$, $M_\mathrm{S}=10$ and various $N$ values, the optimal measurement modes are shown in Fig.~\ref{fig:4}(e). The numerical optimization results for CFI are highly degenerate. We show one set of the optimal measurement modes as an example.

In Fig.~\ref{fig:4}(c), we show the QFI for $\ket{\chi}_{\mathrm{opt}}$ at $D_\mathrm{F}=100$ as red solid line, with the energy allocation into the squeezing portion shown in the inset. The corresponding CFI of an optimal two-mode homodyne receiver is shown as red dotted line denoting the attainable precision of this squeezing enhanced transceiver. In the large total mean photon number regime, this CFI approximately saturates the QFI. For comparison, we show the QFI of a coherent state in its optimal spatial mode, $Z_\mathrm{even}$, as red dashed line, which is saturable by many receivers~\cite{he2024optimum}. The aforesaid squeezing-enhanced transceiver outperforms the optimal classical transceiver in the large total mean photon number regime. 

\begin{figure}
    \centering
\includegraphics[width=\linewidth]{ 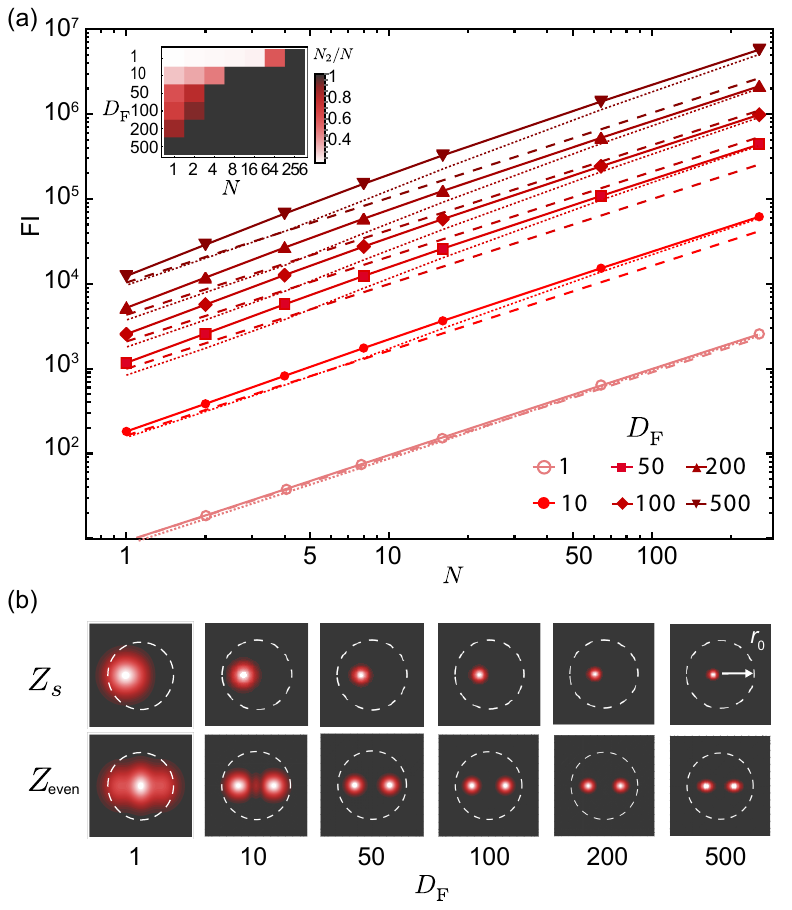}
    \caption{(a) Optimal QFI (solid) for single-mode, squeezing-enhanced probe, optimal QFI (dashed) for coherent-state probe and optimized CFI (dotted) for a two-mode homodyne receiver for different level of diffraction loss ($D_\mathrm{F}$) and mean photon number ($N$). Inset shows the fraction of the total photon energy attributed to the squeezing contribution. (b) Intensity distribution of optimal squeezed, $Z_s$, (assuming N = 256) and coherent, $Z_\mathrm{even}$, transmitter mode for different $D_\mathrm{F}$.}
    \label{fig:5}
\end{figure}

We extend the results presented in Fig.~\ref{fig:4}(c) to different $D_\mathrm{F}$ values in Fig.~\ref{fig:5}(a). The inset of Fig.~\ref{fig:5}(a) shows the fractional energy allocation into the squeezing portion, $N_2/N$, for different $D_\mathrm{F}$ and $N$ values. The CFI attained by the two-mode homodyne receiver, paired with the QFI-optimal single-mode squeezed-state probe (red dotted lines), is seen to saturate the optimal QFI of that probe (red solid lines) when $N>16$, for all the values of $D_\mathrm{F}$ we consider ($1\leq D_\mathrm{F}\leq500$). In the regime where $N>16$, the performance improvement afforded by the squeezing-enhanced transceiver over the optimal classical transceiver (red dashed lines) diminishes with decreasing of $D_\mathrm{F}$. In the far-field regime, $D_\mathrm{F} \leq 1$, only one spatial mode couples any appreciable power over the return-path aperture-limited channel, whose transmissivity is roughly $D_\mathrm{F}$. In this lossy setting, it is no surprise that our optimization reveals that the optimal single-mode Gaussian state approaches the QFI-optimal classical (coherent state) probe in Ref.~\cite{he2024optimum}. In addition to the low-$D_\mathrm{F}$ regime where quantum-enhancement is limited, also in the low total mean photon number regime, viz., $N < 5$, our QFI-optimal squeezed-state transmitter, when paired with our two-mode homodyne receiver, does not outperform the optimal classical transceiver from Ref.~\cite{he2024optimum} in the CFI achieved. The intensity distribution of $Z_s$ and $Z_\mathrm{even}$ are shown in Fig.~\ref{fig:5}(b) for $N=128$ and various $D_\mathrm{F}$ values. We know from~\cite{he2024optimum} (particularly Fig.~2), that the optimal spatial modes for the coherent state are highly degenerate, viz., any linear superposition of $Z_\mathrm{even}$ and $Z_\mathrm{odd}$ is QFI-optimal, that is, $Z_\mathrm{even}$ is an example QFI-optimal mode for a classical laser-light probe. However, the degeneracy is broken in the quantum enhanced case, i.e., $Z_s(x,y)$ and $Z_s(-x,y)$ are uniquely QFI optimal (due to symmetry) for squeezed-state excitation, but not any superposition of the two. This is because of the quantitatively different QFI accrual for phase and transmissivity modulation---of the beam-displacement parameter of interest---on to a squeezed state probe (unlike on a coherent state probe). We also note that the QFI of a single mode squeezing-enhanced Gaussian state probe saturates at a smaller modal support, $M_\mathrm{S}$, compared to the QFI saturation of a coherent state at the same $D_\mathrm{F}$ value that occurs at a higher $M_\mathrm{S}$. This effect is apparent from the intensity distributions of $Z_s$ and $Z_\mathrm{even}$ modes  shown in Fig.~\ref{fig:5}(b), and is consistent the fact that quantum states (such as displaced-squeezed states) are more sensitive to photon loss (caused due to diffraction, here) compared to coherent states, and that higher-$M_{\mathrm S}$ support accommodates higher-order (and thereby, lossier) spatial modes.

We finish this section with a conjecture on the QFI saturating receiver. For single parameter estimation, it is known that the QFI is always saturable~\cite{braunstein1994QFI,oh2019optimal}. Our QFI optimization results show that $|\epsilon_0|^2\gg\epsilon^2-|\epsilon_0|^2$, meaning the parameter-encoding process can be well approximated by a single-mode phase in a multimode lossy passive channel. For a phase encoded in a squeezed thermal state, a non-Gaussian measurement is required to saturate the QFI~\cite{oh2019optimal}. We therefore conjecture that for the estimation of a small transverse displacement, saturating the QFI of the optimal single-mode Gaussian state will necessitate non-Gaussian elements in the receiver (e.g., cubic-phase or self-Kerr interactions), the design of which is outside the scope of this paper.

\section{Conclusion and Outlook}
We explored the use of multi-spatial-mode squeezed light for estimating small beam displacements in an apertured optical system, comparing against the optimal coherent state studied in Ref.~\cite{he2024optimum}. Our analysis showed that single-mode squeezed light always outperforms a single-mode coherent state in the high-photon-number limit, and that a two-mode homodyne detector~\cite{pluchar2025imaging} is an optimal receiver, which can be constructed with structuring the local oscillator~\cite{squeezed_light_pump}.

A key innovation of our study is a method for identifying the optimal single-mode squeezed state in the presence of diffraction loss and finite modal support. Towards this end, we first showed that, in a lossless setting arrived at by artificially curtailing the spatial mode span ($M_\text{S}$) of the probe field to far below the Fresnel-number product ($D_\text{F}$) of the system, by carefully engineering the spatial mode of the probe that is excited in a quantum state (e.g., squeezed state) of light, one can reduce the problem to estimating a parameter embedded in a linear-passive multimode unitary. In this setting, it is known that a single-mode squeezed vacuum, prepared in an appropriate mode within the linear span of the modes on which the unitary acts, can achieve Heisenberg-limited (HL) scaling of the sensitivity with which the parameter can be estimated~\cite{squeezed-light-cam}. The underlying intuitive reason for this is that the unitary rotation at the transmitter results in the parameter of interest manifesting itself as a phase evolution on the single-mode squeezed-state probe, which is known to achieve HL estimation sensitivity. As shown in previous works ~\cite{2023mode_parameter_estimation}, a single-mode quantum-state probe of the standard aperture-function-adapted basis, such as the Hermite-Gauss (HG) modes for soft-Gaussian apertures, does not provide any enhancement over a classical-laser probe. One can also comprehend this precision enhancement as entanglement-based sensing~\cite{zhuang2018distributed}, where the co-propagating orthogonal spatial modes supported by the near-field diffraction-limited system form a network of sensors in an {\em entangled} state (a single-mode squeezed state and $n-1$ vacuum modes, expressed in any $n$-mode linear rotated mode basis is an entangled state), which experiences correlated phases that are collectively determined by the beam deflection, the parameter of interest. 
 
We then considered the problem of finding the optimum single-mode Gaussian (displaced-squeezed) state probe in the lossy case, by removing the $M_{\mathrm S}$ restriction. We considered the perturbative regime of a small beam displacement and found that the beam displacement's modulation of any single-mode probe in any quantum state can be expressed as a three-mode lossy-bosonic parameter-encoding channel. This helped us numerically find the optimal spatial mode in which to excite a squeezed state for QFI-optimal beam-displacement sensing, which we showed---when paired with a two-mode homodyne detection at the receiver---outperforms the optimal laser-light probe from Ref.~\cite{he2024optimum}. Even though there is no HL sensitivity to be had in this lossy setting, the larger is the total photon budget of the transmitter, and deeper in the near field is the imaging system (i.e., with larger $D_{\mathrm F}$), the higher is the Fisher Information improvement afforded by our single-mode squeezed-state probe over the optimum classical laser-light probe.

Many theoretical questions remain open. For example, it is unknown what the QFI-optimal Guassian probe state is for a lossy multimode passive bosonic channel. We conjecture that it is a single-mode squeezed state, in an appropriate mode within the linear span that the channel acts on. Our analysis also does not distinguish how a single-mode squeezed state performs relative to a two-mode probe with one mode in a coherent state and the other in squeezed-vacuum---the latter may achieve similar QFI, but be more loss resilient~\cite{lang2013optimal}. Multi-parameter estimation in this setting is also an open subject.

One natural application of our theory is imaging the vibrations of a nanomechanical resonator, such as the drum modes of a planar membrane~\cite{pluchar2025imaging,Choi2025Quantum}. The scalar beam-displacement problem here becomes a special case of a broader wavefront sensing problem, in which the transverse profile of the probe field is modulated by a phase mask, and the parameter(s) of interest are function(s) of the spatial phase modulation applied by the target or scene. For other settings like near-field biological imaging~\cite{taylor2013_tracking}, the dynamic phase mask---corresponding to the sample being probed---may not be \textit{a priori} known fully, but its statistical properties may be inferred from libraries of previously-illuminated samples. Under these circumstances, one could develop protocols rooted in multiparameter Bayesian estimation~\cite{grace2020josaa,lee2022bayesian,HLscaling_resourcefriendly, boeschoten2025multimodeestimation}, and potentially borrow tools from quantum machine learning~\cite{zhuang2019supervisedlearning}. One could also develop entanglement-enhanced imaging and sensing protocols that effectively jointly estimate independent parameters~\cite{2003-SD-EXP2D}
or parameters corresponding to non-commuting observables~\cite{xia2023multiparameterandnoncommutable,reichert2024heisenberg, 2006-d-t-HG10,2006-d-t-HG10-theory-exp}.

\section*{Acknowledgements}
WH and SG acknowledge Office of Naval Research (ONR) contract numbers N00014-19-1-2189 and N00014-24-1-2627, and Air Force Office of Scientific Research (AFOSR) contract number FA9550-22-1-0180 for co-sponsoring this research. CNG acknowledges funding support from the National Science Foundation, FET, Award No. 2122337. DJW acknowledges support from the National Science Foundation through award number 2239735. WH acknowledges Andrew J. Pizzimenti and Anthony Brady for helpful comments on the manuscript.

\section*{Disclosure}
The authors declare no conflicts of interest.

\section*{Code and Availability}

The code that supports the findings of this study is openly available~\cite{zenodo2025}. The data are available from the authors upon reasonable request.

\setcounter{equation}{0}
\setcounter{figure}{0}

\appendix
\renewcommand{\thefigure}{S.\arabic{figure}}
\setcounter{figure}{0}
\begin{widetext}
\section{Transverse displacement encoding unitary transformation }\label{app:parameter_encoding}
In this section, we derive the transverse displacement encoding unitary transformation and the Hamiltonian in HG bases. When only a limited number of low order modes are accessible, we diagonalize the Hamiltonian and show the corresponding basis. 

For a free-space line-of-sight system enclosed by soft-Gaussian apertures~\cite{shapiro2005ultimate}, we derived the crosstalk matrix induced by a small transverse displacement along $x$-axis in HG modal basis. These results are summarized in the supplementary material of our previous work~\cite{he2024optimum}, specifically Eq. (10). The crosstalk matrix captures the amplitude-crosstalk between each HG modes at the receiver induced by transverse displacement along $x$-axis. In the large $D_\mathrm{F}$ limit, where diffraction losses for low-order HG modes are minimal, the crosstalk matrix for a given number of modes is well approximated by a skew-Hermitian matrix, corresponding to a multimode passive unitary transformation. This low-loss limit is where the quantum resource is expected to be most helpful, with the available enhancement diminishes as the loss increases~\cite{elusive-H-limit}.

The $(m,n)$-th element of the crosstalk matrix is,
\begin{equation}
\begin{aligned}
Q(\tilde{d})_{m,n}&\equiv\iint\displaylimits_{-\infty}^{\infty}\phi_{m}^{\ast}\left(x\right)\phi_{0}^{\ast}\left(y\right)\phi_{n}(x+d) \phi_{0}(y) \times \exp \left( \frac{\left(x+d\right)^2+y^2}{ r_\mathrm{R}^{2}}\right)\exp \left(-\frac{x^{2}+y^2}{ r_\mathrm{R}^{2}}\right)\mathrm{d}x\mathrm{d}y.
\end{aligned}
\end{equation}
Carrying out the integral in the small displacement limit~\cite{he2024optimum}, the crosstalk matrix can be put in the form,
\begin{equation}
\begin{aligned}
Q(\tilde{d})
\approx&\boldsymbol{\mathcal{I}}+\tilde{d}\left[(1+4D_\text{F})^{1/4} \Gamma+\frac{ |\Gamma|}{(1+4D_\text{F})^{\frac{1}{4}}}+\frac{i r_\mathrm{R} 2 \sqrt{D_\text{F}}}{r_\text{T} (1+4D_\text{F})^{\frac{1}{4}}} |\Gamma|\right],\\
\approx &\boldsymbol{\mathcal{I}}+\tilde{d}\left[(1+4D_\text{F})^{\frac{1}{4}} \Gamma+\frac{i r_\text{R} 2 \sqrt{D_\text{F}}}{r_\text{T} (1+4D_\text{F})^{\frac{1}{4}}} |\Gamma|\right],\\
\approx& \boldsymbol{\mathcal{I}}+(4 D_\mathrm{F})^{\frac{1}{4}}\tilde{d} \left[\Gamma+i|\Gamma|\right],
\end{aligned}
\label{eq:crosstalkmatrix}
\end{equation}
where $\boldsymbol{\mathcal{I}}$ is an identity matrix, $\tilde{d}\equiv d/r_\mathrm{R}$ is the normalized transverse displacement, $r_\mathrm{T(R)}$ represents the radius of the transmitter (receiver) aperture, $D_\mathrm{F}$ is the Fresnel number product~\cite{shapiro2005ultimate}, and $\Gamma$ is a tridiagonal skew symmetric matrix~\cite{he2024optimum}, whose $M_\mathrm{S}\times M_\mathrm{S}$ dimensional leading principal submatrix is given by:
\begin{equation}
\Gamma\left(M_{\mathrm{S}}\right)\equiv\left[\begin{array}{ccccc}
0 & 1 & 0 & \ldots & 0 \\
-1 & 0 & \sqrt{2} & \ldots & 0 \\
0 & -\sqrt{2} & 0 & \ddots & \vdots \\
\vdots & \vdots & \ddots & \ddots & \sqrt{M_{\mathrm{S}}-1} \\
0 & 0 & \ldots & -\sqrt{M_{\mathrm{S}}-1} & 0
\end{array}\right].
\end{equation}
The approximation in the second row of Eq.~\eqref{eq:crosstalkmatrix} holds when $D_\mathrm{F}\gg1$ limit. In the first row of Eq.~\eqref{eq:crosstalkmatrix}, the third term is smaller than the second and fourth terms by a factor, $(1+4 D_\text{F})^{1/2}\approx 2 \sqrt{D_\text{F}}$. When $D_\mathrm{F}$ is on the order of 100, this ratio is more than 20. The approximation in the third row of Eq.~\eqref{eq:crosstalkmatrix} holds in the limit where $D_\mathrm{F}\gg1$ and we take $r_\mathrm{T}=r_\mathrm{R}=r_0$. Under these conditions, we denote the crosstalk matrix for the first $M_\mathrm{S}+1$ modes by,
\begin{equation}
    Q_1(\tilde{d})\equiv \boldsymbol{\mathcal{I}}(M_\mathrm{S}+1)+(4 D_\mathrm{F})^{\frac{1}{4}}\tilde{d} \left[\Gamma(M_\mathrm{S}+1)+i|\Gamma(M_\mathrm{S}+1)|\right].    
\end{equation}
This is a skew Hermitian matrix. The corresponding unitary operator is
\begin{equation}\label{eq:HG_U}
    \hat{U}(\tilde{d})=\exp{-\vec{\hat{a}}^\dagger \ln{\left(Q_1\right)}\vec{\hat{a}}}\equiv\exp{-i \tilde{d} \hat{H}}.
\end{equation}
The Hamiltonian is parameter independent,
\begin{equation}
\begin{aligned}
    \hat{H}\approx&-i (4 D_\mathrm{F})^{\frac{1}{4}}\vec{\hat{a}}^\dagger \left(\Gamma+i|\Gamma|\right)\vec{\hat{a}},\\
    =& 2 D_\mathrm{F}^{\frac{1}{4}} \sum\displaylimits_{n=0}^{M_\mathrm{S}-1} \sqrt{n+1} \left(e^{-\frac{i\pi}{4}}\hat{a}_{n+1}^{}\hat{a}_{n}^\dagger+e^{\frac{i\pi}{4}}\hat{a}_{n}^{}\hat{a}_{n+1}^\dagger
    \right).
    \label{eq:Hamiltonian_HG}
\end{aligned}
\end{equation}
where $\vec{\hat{a}}$ represents a column vector comprising of lowering operators $\hat{a}_n$ corresponding to the receiver-side HG modes $\phi_n(x)\phi_0(y)$. Therefore, the parameter encoding transformation represented in the HG mode basis is composed of consecutive, pair-wise beam splitters as shown in the green dashed box of Fig.~\ref{fig:2} (a). 

Due to the fact that the Hamiltonian is parameter-independent, performing proper transverse displacement independent modal transformation on the first $M_\mathrm{S}$ modes, $\hat{U}
(\tilde{d})$ can be decomposed into a series of parameter-dependent single-mode phases, followed by beam splitter interactions with mode $\hat{a}_{M_\mathrm{S}}$, which is assumed to always starts in vacuum. This modal transformation can be found by diagonalizing the crosstalk matrix. Since $\Gamma$ is a real skew symmetric matrix, we can diagonalize it. For an $M_\mathrm{S}$ mode transmitter, we have
\begin{equation}
\begin{aligned}   
&\boldsymbol{\mathcal{I}}(M_\mathrm{S}) +\left(4 D_\mathrm{F}\right)^{\frac{1}{4}} \tilde{d} \left[B T R\left(\Gamma\left(M_\mathrm{S}\right)+i |\Gamma\left(M_\mathrm{S}\right)|\right)R^{\dagger} T^{T} B^{\dagger}\right] =  \bigoplus_{n=1}^{ M_{S}/2 }
 \left(\begin{array}{cc}
1+2 i \lambda_n D_\mathrm{F}^{\frac{1}{4}} \tilde{d}     &  0\\
  0    & 1-2 i \lambda_n D_\mathrm{F}^{\frac{1}{4}} \tilde{d} 
 \end{array}\right) ,
    \label{eq:rotation}
\end{aligned}
\end{equation}
in which $R$ is a diagonal matrix with single-mode phases, $R_{n,n}\equiv\exp{\frac{i \pi}{4} (n-1)}$.  The matrix $B$ represents the direct sum of $M_\mathrm{S}/2$ matrices of pair-wise $50:50$ beam splitter,
$    B\equiv\bigoplus_{n=1}^{ M_\mathrm{S}/2} B_0$, where we have,
\begin{equation}
    B_0\equiv
    \frac{1}{\sqrt{2}}\left(\begin{array}{cc}
       i  &  1\\
       -i  & 1
    \end{array}\right).
\end{equation}
The matrix $T$ represents the orthogonal transformation that brings $\Gamma(M_\mathrm{S})$ into a block-diagonal form. 
\begin{equation}\label{eq:def_lambda}
T\Gamma(M_\mathrm{S})T^\mathrm{T}=\bigoplus_{n=1}^{ M_\mathrm{S}/2 }
 \left(\begin{array}{cc}
0& i \lambda_n    \\
   - i \lambda_n & 0   
 \end{array}\right).
\end{equation}
An orthogonal transformation is realizable with a simple passive unitary transformation.  The eigenvalues of $\Gamma$ are all imaginary, represented as $\pm i\lambda_n$. We order $\lambda_n$ in a non-increasing order.

With the modal transformation defined above, we find a new modal basis 
\begin{equation}\label{eq:single_phase_basis}
    \vec{\hat{b}}\equiv B T R \vec{\hat{a}}.
\end{equation}

We can represent the Hamiltonian in terms of $\hat{b}$,
\begin{equation}
\begin{aligned}
\hat{H}=&\sum \displaylimits_{n=1}^{M_\mathrm{S}/2} -2  \lambda_n D_\mathrm{F}^{\frac{1}{4}} \left(\hat{b}_{2n-2}^\dagger\hat{b}_{2n-2}-\hat{b}_{2n-1}^\dagger\hat{b}_{2n-1}\right)+2 D_\mathrm{F}^{\frac{1}{4}} \sqrt{M_\mathrm{S}}\left(e^{-\frac{i \pi}{4}} \hat{a}_{M_\mathrm{S}}\hat{a}_{M_\mathrm{S}-1}^\dagger+e^{\frac{i \pi}{4}} \hat{a}_{M_\mathrm{S}-1}\hat{a}_{M_\mathrm{S}}^\dagger\right).
\end{aligned}\label{eq:supp_Hb1}
\end{equation}
\begin{figure}
    \centering
\includegraphics[width=.5\linewidth]{ 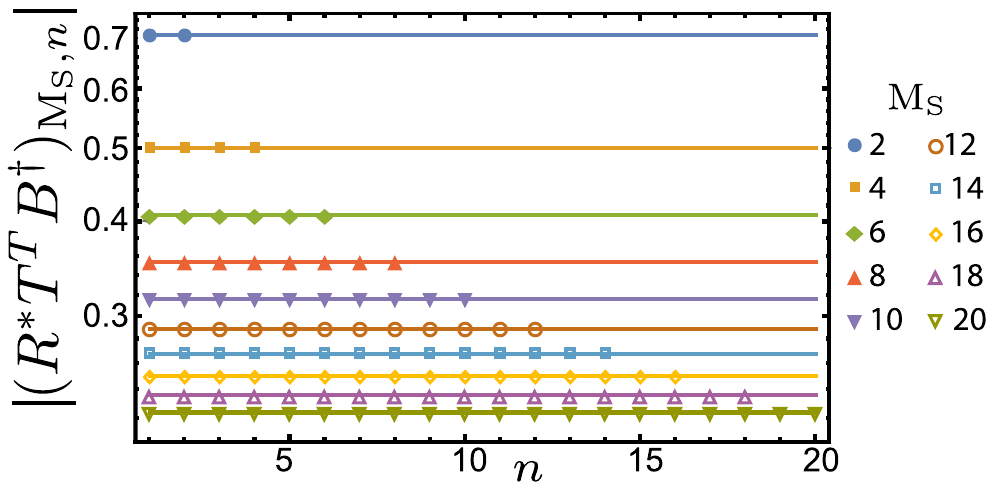}
    \caption{The absolute value of the last row of the transformation matrix $R^\ast T^T B^\dagger$ is plotted, along with the gridlines for 
$1/\sqrt{M_\mathrm{S}}$ using the same color.}
    \label{fig:lastrow}
\end{figure}
Now we also replace $\hat{a}_{M_\mathrm{S}-1}$. Based on numerically solving the transformation matrix, $R^\ast T^T B^\dagger$, and examine the last row (shown in Fig.~\ref{fig:lastrow}), we \textbf{conjecture} that
\begin{equation}
\begin{aligned}\label{eq:ams_to_b}
    \hat{a}_{M_\mathrm{S}-1}=& \left(R^\ast T^T B^\dagger\right)_{M_\mathrm{S}} \cdot\vec{\hat{b}}
    = \frac{1}{\sqrt{M_\mathrm{S}}}\sum\displaylimits_{n=0}^{M_\mathrm{S}-1} e ^{i\varrho_n}\hat{b}_n,
\end{aligned}
\end{equation} 
where $\varrho_n$ represents some phase factor. 
By substituting $\hat{a}_{M_\mathrm{S}-1}$ into Eq.~\eqref{eq:supp_Hb1}, we get
\begin{equation}
\begin{aligned}
    \hat{H}=&\sum \displaylimits_{n=1}^{M_\mathrm{S}/2} -2  \lambda_n D_\mathrm{F}^{\frac{1}{4}} \left(\hat{b}_{2n-2}^\dagger\hat{b}_{2n-2}-\hat{b}_{2n-1}^\dagger\hat{b}_{2n-1}\right)+2 D_\mathrm{F}^{\frac{1}{4}}\sum\displaylimits_{n=0}^{M_\mathrm{S}-1}\left(e^{-\frac{i \pi}{4}} \hat{a}_{M_\mathrm{S}}\hat{b}_{n}^\dagger e ^{-i\varrho_n}+e^{\frac{i \pi}{4}} e ^{i\varrho_n}\hat{b}_{n}\hat{a}_{M_\mathrm{S}}^\dagger\right),
\end{aligned}
\label{eq:Hb}
\end{equation}
 The Hamiltonian in Eq.~\eqref{eq:Hb} indicates that, the transverse-displacement-induced parameter encoding can be represented as a strength-decreasing single-mode phase evolution on each $\hat{b}$ mode, followed by an equal strength beam splitter with $\hat{a}_{M_\mathrm{S}}$ which is in vacuum state at the transmitter. 
 The strength of the single-mode phase evolution is denoted by $2D_\mathrm{F}^{\frac{1}{4}}\lambda_n$. We also note an important property: the new spatial modes, $\hat{b}$, no longer have well-defined spatial mode parity as the HG modes do, meaning they are not strictly even or odd functions.

\section{Evaluation of QFI} \label{app:K_calc}
Firstly, we evaluate the QFI for a single-mode displaced-squeezed state of mode $\mathrm{HG}_{n,0}$.  
\begin{equation}
\begin{aligned}
K\left[|0\rangle |\boldsymbol{\alpha};r e^{i\vartheta_1}\rangle |0\rangle,\hat{H}_1\right] \equiv& 4 \left<\Delta^2\hat{H}_1 \right>_{|0\rangle |\boldsymbol{\alpha};r e^{i \vartheta_1}\rangle |0\rangle},\\
    =&4 \left(\bra{0}\bra{\boldsymbol{\alpha};r e^{i \vartheta_1}}\bra{0}\hat{H}_1^2 \ket{0}\ket{\boldsymbol{\alpha};r e^{i \vartheta_1}}\ket{0}-\bra{0}\bra{\boldsymbol{\alpha};r e^{i \vartheta_1}}\bra{0}\hat{H}_1 \ket{0}\ket{\boldsymbol{\alpha};r e^{i \vartheta_1}}\ket{0 }^2\right).
    \label{eq:k1supp}
\end{aligned}
\end{equation}
The Hamiltonian $\hat{H}_1$ involves the two adjacent modes, $\mathrm{HG}_{n\pm1,0}$, each of which is in a vacuum state. Thus we can write this single-mode excited state as $\ket{0}_{n-1}\ket{\boldsymbol{\alpha};r e^{i \vartheta_1} }_{n}\ket{0}_{n+1}$. Below we will omit the three subscripts indicating the corresponding HG mode order for notation simplicity. We evaluate the two terms in the last row of Eq.~\eqref{eq:k1supp} separately.

The first term is,
\begin{equation}\begin{aligned}
&\bra{0}\bra{\boldsymbol{\alpha};r e^{i \vartheta_1} }\bra{0} \hat{H}_1^2 \ket{0}\ket{\boldsymbol{\alpha};r e^{i \vartheta_1} }\ket{0 }\\=&\bra{0}\bra{\boldsymbol{\alpha};r e^{i \vartheta_1}}\bra{0} 4 D_\mathrm{F}^{\frac{1}{2}} \left\{ \sqrt{n} \left(e^{-\frac{i\pi}{4}}\hat{a}_{n}\hat{a}_{n-1}^\dagger+e^{\frac{i\pi}{4}}\hat{a}_{n-1}\hat{a}_{n}^\dagger
    \right)+\sqrt{n+1} \left(e^{-\frac{i\pi}{4}}\hat{a}_{n+1}\hat{a}_{n}^\dagger+e^{\frac{i\pi}{4}}\hat{a}_{n}\hat{a}_{n+1}^\dagger
    \right) \right\}^2\ket{0}\ket{\boldsymbol{\alpha};r  e^{i \vartheta_1}}\ket{0 },\\
    =& 4 \sqrt{D_\mathrm{F}}\bra{0}\bra{\boldsymbol{\alpha};r e^{i \vartheta_1}}\bra{0} n \left(e^{-\frac{i\pi}{4}}\hat{a}_{n}\hat{a}_{n-1}^\dagger+e^{\frac{i\pi}{4}}\hat{a}_{n-1}\hat{a}_{n}^\dagger
    \right)^2+(n+1) \left(e^{-\frac{i\pi}{4}}\hat{a}_{n+1}\hat{a}_{n}^\dagger+e^{\frac{i\pi}{4}}\hat{a}_{n}\hat{a}_{n+1}^\dagger
    \right)^2 \ket{0}\ket{\boldsymbol{\alpha};r  e^{i \vartheta_1}}\ket{0 }\\
    +&  4 \sqrt{D_\mathrm{F}}\bra{0}\bra{\boldsymbol{\alpha};r e^{i \vartheta_1}}\bra{0} \sqrt{n(n+1)} \left(e^{-\frac{i\pi}{4}}\hat{a}_{n}\hat{a}_{n-1}^\dagger+e^{\frac{i\pi}{4}}\hat{a}_{n-1}\hat{a}_{n}^\dagger
    \right)\left(e^{-\frac{i\pi}{4}}\hat{a}_{n+1}\hat{a}_{n}^\dagger+e^{\frac{i\pi}{4}}\hat{a}_{n}\hat{a}_{n+1}^\dagger
    \right) \ket{0}\ket{\boldsymbol{\alpha};r e^{i \vartheta_1}}\ket{0}\\
    +&  4 \sqrt{D_\mathrm{F}}\bra{0}\bra{\boldsymbol{\alpha};r e^{i \vartheta_1}}\bra{0}\sqrt{n(n+1)} \left(e^{-\frac{i\pi}{4}}\hat{a}_{n+1}\hat{a}_{n}^\dagger+e^{\frac{i\pi}{4}}\hat{a}_{n}\hat{a}_{n+1}^\dagger
    \right) \left(e^{-\frac{i\pi}{4}}\hat{a}_{n}\hat{a}_{n-1}^\dagger+e^{\frac{i\pi}{4}}\hat{a}_{n-1}\hat{a}_{n}^\dagger
    \right)\ket{0}\ket{\boldsymbol{\alpha};r e^{i \vartheta_1}}\ket{0 }.
\end{aligned}
\end{equation}
We know that,
\begin{equation}\begin{aligned}
   \prescript{}{j}{\bra{0}}\hat{a}_{j}^\dagger\ket{0}_j&=
\prescript{}{j}{\bra{0}}\hat{a}_{j}\ket{0}_j=0,\\
\prescript{}{j}{\bra{0}}(\hat{a}_{j}^\dagger)^2\ket{0}_j&=
\prescript{}{j}{\bra{0}}(\hat{a}_{j})^2\ket{0}_j=0,
\label{eq:simplify}
\end{aligned}
\end{equation}
where $j=n\pm1$.
Hence the first term is simplified:
\begin{equation}
    \begin{aligned}
&\bra{0}\bra{\boldsymbol{\alpha};r e^{i \vartheta_1}}\bra{0} \hat{H}_1^2 \ket{0}\ket{\boldsymbol{\alpha};r e^{i \vartheta_1}}\ket{0 }\\
=&4 \sqrt{D_\mathrm{F}}\bra{0}\bra{\boldsymbol{\alpha};r e^{i \vartheta_1}}\bra{0} n \left(\hat{a}_{n}\hat{a}_{n-1}^\dagger \hat{a}_{n-1}\hat{a}_{n}^\dagger+\hat{a}_{n-1}\hat{a}_{n}^\dagger\hat{a}_{n}\hat{a}_{n-1}^\dagger\right) \ket{0}\ket{\boldsymbol{\alpha};r e^{i \vartheta_1}}\ket{0 },\\
&+4 \sqrt{D_\mathrm{F}}\bra{0}\bra{\boldsymbol{\alpha};r e^{i \vartheta_1}}\bra{0} (n+1) \left(\hat{a}_{n+1}\hat{a}_{n}^\dagger \hat{a}_{n}\hat{a}_{n+1}^\dagger+\hat{a}_{n}\hat{a}_{n+1}^\dagger\hat{a}_{n+1}\hat{a}_{n}^\dagger\right) \ket{0}\ket{\boldsymbol{\alpha};r e^{i \vartheta_1}}\ket{0 },\\
=&4 \sqrt{D_\mathrm{F}}\left(\bra{0}\bra{\boldsymbol{\alpha};r e^{i \vartheta_1}}\bra{0} n \hat{a}_{n-1}\hat{a}_{n}^\dagger\hat{a}_{n}\hat{a}_{n-1}^\dagger\ket{0}\ket{\boldsymbol{\alpha};r e^{i \vartheta_1}}\ket{0 }+\bra{0}\bra{\boldsymbol{\alpha};r e^{i \vartheta_1}}\bra{0} (n+1)\hat{a}_{n+1}\hat{a}_{n}^\dagger \hat{a}_{n}\hat{a}_{n+1}^\dagger \ket{0}\ket{\boldsymbol{\alpha};r e^{i \vartheta_1}}\ket{0 }\right),\\
=&4 \sqrt{D_\mathrm{F}}\left(\bra{\boldsymbol{\alpha};r e^{i \vartheta_1}} n \hat{a}_{n}^\dagger\hat{a}_{n}\ket{\boldsymbol{\alpha};r e^{i \vartheta_1}}+\bra{\boldsymbol{\alpha};r e^{i \vartheta_1}} \left(n+1\right)\hat{a}_{n}^\dagger \hat{a}_{n} \ket{\boldsymbol{\alpha};r e^{i \vartheta_1}}\right),\\
=&4 \sqrt{D_\mathrm{F}}N\left(2n+1\right).
\end{aligned}
\label{eq:term1k1}
\end{equation}
The second term is
\begin{equation}\begin{aligned}
&\bra{0}\bra{\boldsymbol{\alpha};r e^{i \vartheta_1}}\bra{0} \hat{H}_1 \ket{0}\ket{\boldsymbol{\alpha};r e^{i \vartheta_1}}\ket{0 }^2\\=&\bra{0}\bra{\boldsymbol{\alpha};r e^{i \vartheta_1}}\bra{0} 2 D_\mathrm{F}^{\frac{1}{4}} \left\{ \sqrt{n} \left(e^{-\frac{i\pi}{4}}\hat{a}_{n}\hat{a}_{n-1}^\dagger+e^{\frac{i\pi}{4}}\hat{a}_{n-1}\hat{a}_{n}^\dagger
    \right)+\sqrt{n+1} \left(e^{-\frac{i\pi}{4}}\hat{a}_{n+1}\hat{a}_{n}^\dagger+e^{\frac{i\pi}{4}}\hat{a}_{n}\hat{a}_{n+1}^\dagger
    \right) \right\}\ket{0}\ket{\boldsymbol{\alpha};r e^{i \vartheta_1}}\ket{0 }^2,\\
    =&0.
    \label{eq:term2k1}
\end{aligned}
\end{equation}
Combining Eq.~\eqref{eq:term1k1}, Eq.~\eqref{eq:term2k1} and Eq.~\eqref{eq:k1supp}, we arrive at Eq.~\eqref{eq:K1}.
 
Next, we evaluate the QFI attained by a single-mode squeezed vacuum state of mode $\hat{b}_0$ given in Eq.~\eqref{eq:b1qfi}. The evolution Hamiltonian is given in Eq.~\eqref{eq:H_bmode}. The other mode involved in $\hat{H}_2$ is $\hat{b}_{M_\mathrm{S}}$ in a vacuum state. The QFI of a pure state is:
\begin{equation}
K\left[\ket{0;r e^{i \vartheta_1}}\ket{0},\hat{H}_2\right]\equiv 4 \left<\Delta^2\hat{H}_2 \right>_{\ket{0;r e^{i \vartheta_1}}\ket{0}}=4\left(\bra{0;r e^{i \vartheta_1}}\bra{0}\hat{H}_2^2\ket{0;r e^{i \vartheta_1}}\ket{0}-\bra{0;r e^{i \vartheta_1}}\bra{0}\hat{H_2}\ket{0;r e^{i \vartheta_1}}\ket{0}^2\right).
\end{equation}
The first term is:
\begin{equation}
    \begin{aligned}
&\bra{0;r e^{i \vartheta_1}}\bra{0}\hat{H}_2^2\ket{0;r e^{i \vartheta_1}}\ket{0}\\
=&\bra{0;r e^{i \vartheta_1}}\bra{0}\left\{-2  \lambda_1 D_\mathrm{F}^{\frac{1}{4}} \hat{b}_{0}^\dagger\hat{b}_{0}+2 D_\mathrm{F}^{\frac{1}{4}}\left(\hat{b}_{M_\mathrm{S}}\hat{b}_{0}^\dagger+\hat{b}_{0}\hat{b}_{M_\mathrm{S}}^\dagger\right)\right\}^2\ket{0;r e^{i \vartheta_1}}\ket{0},\\
=&\bra{0;r e^{i \vartheta_1}}\bra{0}
4  \lambda_1^2 D_\mathrm{F}^{\frac{1}{2}} \hat{b}_{0}^\dagger\hat{b}_{0}\hat{b}_{0}^\dagger\hat{b}_{0}+4 D_\mathrm{F}^{\frac{1}{2}}\left(\hat{b}_{M_\mathrm{S}}\hat{b}_{0}^\dagger+\hat{b}_{0}\hat{b}_{M_\mathrm{S}}^\dagger\right)^2-4  \lambda_1 D_\mathrm{F}^{\frac{1}{2}} \hat{b}_{0}^\dagger\hat{b}_{0} \left(\hat{b}_{M_\mathrm{S}}\hat{b}_{0}^\dagger+\hat{b}_{0}\hat{b}_{M_\mathrm{S}}^\dagger\right)\ket{0;r e^{i \vartheta_1}}\ket{0}\\
-&\bra{0;r e^{i \vartheta_1}}\bra{0}4 \lambda_1 D_\mathrm{F}^{\frac{1}{2}}\left(\hat{b}_{M_\mathrm{S}}\hat{b}_{0}^\dagger+\hat{b}_{0}\hat{b}_{M_\mathrm{S}}^\dagger\right)\hat{b}_{0}^\dagger\hat{b}_{0}
\ket{0;r e^{i \vartheta_1}}\ket{0},\\
\end{aligned}\label{eq:k2_h2square}
\end{equation}
Leveraging the fact that $\hat{b}_{M_\mathrm{S}}$ mode is in a vacuum state, we have $\bra{0}\hat{b}_{M_\mathrm{S}}^\dagger\ket{0}=
\bra{0}\hat{b}_{M_\mathrm{S}}\ket{0}=0$ and $\bra{0}(\hat{b}_{M_\mathrm{S}}^\dagger)^2\ket{0}=
\bra{0}(\hat{b}_{M_\mathrm{S}})^2\ket{0}=0$. We can simplify Eq.~\eqref{eq:k2_h2square} into:
\begin{equation}
     \begin{aligned}
&\bra{0;r  e^{i \vartheta_1}}\bra{0}\hat{H}_2^2\ket{0;r  e^{i \vartheta_1}}\ket{0}\\
=&\bra{0;r  e^{i \vartheta_1}}\bra{0}
4  \lambda_1^2 D_\mathrm{F}^{\frac{1}{2}} \hat{b}_{0}^\dagger\hat{b}_{0}^{}\hat{b}_{0}^\dagger\hat{b}_{0}^{}+4 D_\mathrm{F}^{\frac{1}{2}}\left(\hat{b}_{M_\mathrm{S}}^{}\hat{b}_{0}^\dagger+\hat{b}_{0}^{}\hat{b}_{M_\mathrm{S}}^\dagger\right)^2\ket{0;r  e^{i \vartheta_1}}\ket{0},\\
=&\bra{0;r  e^{i \vartheta_1}}
4  \lambda_1^2 D_\mathrm{F}^{\frac{1}{2}} \hat{b}_{0}^\dagger\hat{b}_{0}^{}\hat{b}_{0}^\dagger\hat{b}_{0}^{}\ket{0;r  e^{i \vartheta_1}}+4 D_\mathrm{F}^{\frac{1}{2}}\bra{0;r  e^{i \vartheta_1}}\bra{0}
\hat{b}_{M_\mathrm{S}}^{}\hat{b}_{0}^\dagger\hat{b}_{0}^{}\hat{b}_{M_\mathrm{S}}^\dagger+\hat{b}_{0}^{}\hat{b}_{M_\mathrm{S}}^\dagger\hat{b}_{M_\mathrm{S}}^{}\hat{b}_{0}^\dagger\ket{0;r  e^{i \vartheta_1}}\ket{0},\\
=&4  \lambda_1^2 D_\mathrm{F}^{\frac{1}{2}} \bra{0;r  e^{i \vartheta_1}}\hat{b}_{0}^\dagger\hat{b}_{0}^{}\hat{b}_{0}^\dagger\hat{b}_{0}^{}\ket{0;r  e^{i \vartheta_1}}+4 D_\mathrm{F}^{\frac{1}{2}}\bra{0;r  e^{i \vartheta_1}}
\hat{b}_{0}^\dagger\hat{b}_{0}^{}\ket{0;r  e^{i \vartheta_1}}.
\end{aligned}
\end{equation}
The two terms in the last row are the first and second order moments of the number operator, $\hat{N}_{b_0}\equiv\hat{b}_0^\dagger\hat{b}_0^{}$ for a squeezed vacuum state. By applying a squeezing operator on a vaccum state we get a squeezed vacuum state, $\ket{0;r  e^{i \vartheta_1}}= \hat{S}(r e^{i \vartheta_1})\ket{0}\equiv\exp\left(\frac{r}{2} \left(e^{i \vartheta_1} \hat{b}_0^{\dagger2}-e^{-i \vartheta_1}\hat{b}_0^2\right)\right)\ket{0}$. We can write out the field operator transformed by the squeezing operator~\cite{ferraro2005gaussianstatereview}, 
\begin{equation}
    \begin{aligned}
\hat{S}^\dagger\left(r e^{i \vartheta_1}\right)\hat{b}_0\hat{S}\left(r e^{i \vartheta_1}\right)=&\hat{b}_0 \mathrm{cosh}\left(r\right)+\hat{b}_0^\dagger \mathrm{sinh}(r) e^{i\vartheta_1},\\ 
\hat{S}^\dagger\left(r e^{i \vartheta_1} \right)\hat{b}_0^\dagger\hat{S}\left(r e^{i \vartheta_1}\right)=&\hat{b}_0^\dagger \mathrm{cosh}(r)+\hat{b}_0\mathrm{sinh}(r) e^{-i\vartheta_1}.
    \end{aligned}
\end{equation}
It is straightforward to obtain:
\begin{equation}
\begin{aligned}
\bra{0;r  e^{i \vartheta_1}}\hat{b}_{0}^\dagger\hat{b}_{0}^{}\ket{0;r  e^{i \vartheta_1}}&=\bra{0}\hat{S}^\dagger\left(r e^{i\vartheta_1}\right)\hat{b}_{0}^\dagger\hat{S}\left(r e^{i\vartheta_1}\right)\hat{S}^\dagger\left(r e^{i\vartheta_1}\right)\hat{b}_{0}^{}\hat{S}\left(r e^{i\vartheta_1}\right)\ket{0},\\
&=\bra{0}\left(\hat{b}_0^\dagger \mathrm{cosh}(r)+\hat{b}_0\mathrm{sinh}(r) e^{-i\vartheta_1}\right)
\left(\hat{b}_0 \mathrm{cosh}(r)+\hat{b}_0^\dagger \mathrm{sinh}(r) e^{i\vartheta_1}\right)\ket{0},\\
&=\bra{0}\mathrm{sinh}^2(r)\hat{b}_0\hat{b}_0^\dagger\ket{0}=\mathrm{sinh}^2(r)=N.\\
\end{aligned}
\end{equation}
Applying the same rules, we have:
\begin{equation}
\bra{0;r  e^{i \vartheta_1}}\hat{b}_{0}^\dagger\hat{b}_{0}^{}\hat{b}_{0}^\dagger\hat{b}_{0}^{}\ket{0;r  e^{i \vartheta_1}}=3N^2+2N.
\end{equation}
Thus we arrive at,
\begin{equation}
    \bra{0;r  e^{i \vartheta_1}}\hat{H}_2^2\ket{0;r  e^{i \vartheta_1}}=4  \lambda_1^2 D_\mathrm{F}^{\frac{1}{2}} (3N^2+2N)+4 D_\mathrm{F}^{\frac{1}{2}} N.
\end{equation}
Then we evaluate the second term,
\begin{equation}
\begin{aligned}
\bra{0;r  e^{i \vartheta_1}}\bra{0}\hat{H}_2\ket{0;r  e^{i \vartheta_1}}\ket{0}^2=&\bra{0;r  e^{i \vartheta_1}}\bra{0}-2  \lambda_1 D_\mathrm{F}^{\frac{1}{4}} \hat{b}_{0}^\dagger\hat{b}_{0}+2 D_\mathrm{F}^{\frac{1}{4}}\left(\hat{b}_{M_\mathrm{S}}\hat{b}_{0}^\dagger+\hat{b}_{0}\hat{b}_{M_\mathrm{S}}^\dagger\right)\ket{0;r  e^{i \vartheta_1}}\ket{0}^2,\\
=&\bra{0;r  e^{i \vartheta_1}}-2  \lambda_1 D_\mathrm{F}^{\frac{1}{4}} \hat{b}_{0}^\dagger\hat{b}_{0}\ket{0;r  e^{i \vartheta_1}}^2,\\
=&4 \lambda_1^2D_\mathrm{F}^{\frac{1}{2}} N^2.
\end{aligned}
\end{equation}
Combining the results above together we have,
\begin{equation}\label{eq:k2}
    K\left[\ket{0;r e^{i \vartheta_1}}\ket{0},\hat{H}_2\right]=4 D_\mathrm{F}^{\frac{1}{2}}\left[ 4 \lambda_1^2(3N^2+2N)+4 N-4 \lambda_1^2 N^2\right]=32D_\mathrm{F}^{\frac{1}{2}}\lambda_1^2 N(N+1)+16 D_\mathrm{F}^{\frac{1}{2}}N.
\end{equation}

\section{CFI Evaluation of a two-mode homodyne receiver}\label{app:Lossless_twomode_homodyne}
For a single-mode squeezed-vacuum-state probe of mode $\hat{b}_0$, the QFI is given in Eq.~\eqref{eq:b1qfi} and repeated in Eq.~\eqref{eq:k2}. We show a QFI saturating receiver in this appendix. The relevant parameter-encoding Hamiltonian is $\hat{H}_2$ and the corresponding unitary transformation is $\hat{U}_2\equiv\exp\{-i \tilde{d} \hat{H}_2\}$. We show a two-mode homodyne measurement can saturate the QFI where the measurement is performed in mode $\hat{b}_0^\prime\equiv\hat{U}_2(\tilde{d})^\dagger\hat{b}_0\hat{U}_2(\tilde{d})$ and $\hat{b}_{M_\mathrm{S}}^\prime\equiv\hat{U}_2(\tilde{d})^\dagger\hat{b}_{M_\mathrm{S}}\hat{U}_2(\tilde{d})$ that samples the quadrature along $\mathrm{ArcTan}(e^{2r})$ and $\frac{\pi}{4}$ respectively, as shown in Fig.~\ref{fig:3}(a). In the analysis below, we use the mean vector and covariance matrix to represent a Gaussian state, and we use the symplectic matrix to represent a Gaussian unitary transformation. The symplectic matrix can act on the mean vector and covariance matrix to represent a state evolving through a transformation~\cite{weedbrook2012gaussian}. 

We first introduce the mean vector and covariance matrix representation of a Gaussian state. For a certain mode, $\hat{b}_j$, following the formalism in~\cite{weedbrook2012gaussian}, the \textit{quadrature field operators} are $\hat{q}_j=\hat{b}_j^{}+\hat{b}_j^\dagger$ and $\hat{p}_j=-i (\hat{b}_j^{}+\hat{b}_j^\dagger)$, where $i$ represents the imaginary unit, and $j$ represents an integer ranging from all the modes we keep track of. For an $M$ mode state, we group the quadrature field operators into a column vector $\hat{X}=\{\hat{q}_1,\hat{q}_2,\cdots, \hat{q}_{M},\hat{p}_1,\hat{p}_2,\cdots\hat{p}_M\}^{T}$. A Gaussian state can be fully characterized by its mean vector and covariance matrix. The $j$-th element of the mean vector is defined as 
\begin{equation}
    \begin{aligned}
X_{\{j\}}&\equiv\left<\hat{X}_j\right>.\\
    \end{aligned}
\end{equation}
The $\{j,l\}$-th element of the covariance matrix is given by
\begin{equation}
    \begin{aligned}
V_{\{j,l\}}&\equiv \frac{1}{2}\left<(\hat{X}_j-X_j)(\hat{X}_l-X_l)+(\hat{X}_l-X_l)(\hat{X}_j-X_j) \right>.
    \end{aligned}
\end{equation}
In the above definitions $j,l\in\{1,2,\cdots,M\}$. A two-mode state that is a tensor product of a squeezed vacuum and a vacuum state can be presented as, $\ket{0;r}\ket{0}=\exp(\frac{r}{2} ( \hat{b}_0^{\dagger2}-\hat{b}_0^2))\ket{0}\ket{0}$~\cite{ferraro2005gaussianstatereview}. We can see the QFI in Eq.~\eqref{eq:b1qfi} is independent of the squeezed vacuum state's squeezing angle $\vartheta_1$, thus without loss of generality we set $\vartheta_1=0$. The corresponding mean vector and covariance matrix is,
\begin{equation}
    \begin{aligned}
        X=&\{0,0,0,0\}^T,\\
        V=&\left(\begin{array}{cccc}
          e^{2r}  , & 0 ,&0,&0\\
           0,&1,&0,&0\\
           0,   &  0,& e^{-2r},&0\\   0,&0,&0,&1
        \end{array}\right).
    \end{aligned}
\end{equation}

Next we show the unitary matrix representation of $\hat{U}_2(\tilde{d})$. Then, we show the corresponding symplectic matrix representation. The parameter encoding unitary can be expanded,
\begin{equation}
\begin{aligned}
    \hat{U}_2(\tilde{d})&\equiv \exp\{-i \tilde{d} \hat{H}_2\},\\
    &=\exp\left\{-i \tilde{d} \left(-2  \lambda_1 D_\mathrm{F}^{\frac{1}{4}} \hat{b}_{0}^\dagger\hat{b}_{0}+2 D_\mathrm{F}^{\frac{1}{4}}\left(\hat{b}_{M_\mathrm{S}}\hat{b}_{0}^\dagger+\hat{b}_{0}\hat{b}_{M_\mathrm{S}}^\dagger\right)\right) \right\},\\
    &=\exp\left\{-i 2 \tilde{d} D_\mathrm{F}^{\frac{1}{4}}\left(\hat{b}_{M_\mathrm{S}}\hat{b}_{0}^\dagger+\hat{b}_{0}\hat{b}_{M_\mathrm{S}}^\dagger\right)\right\}\exp\left\{i 2 \tilde{d}   \lambda_1 D_\mathrm{F}^{\frac{1}{4}} \hat{b}_{0}^\dagger\hat{b}_{0}\right\} \exp\left\{\frac{-(\tilde{d} 2 D_\mathrm{F}^{\frac{1}{4}})^2}{2}[\left(\hat{b}_{M_\mathrm{S}}\hat{b}_{0}^\dagger+\hat{b}_{0}\hat{b}_{M_\mathrm{S}}^\dagger\right),  \lambda_1\hat{b}_{0}^\dagger\hat{b}_{0}]\right\}\cdots,\\
    &\approx\exp\left\{-i 2 \tilde{d} D_\mathrm{F}^{\frac{1}{4}}\left(\hat{b}_{M_\mathrm{S}}\hat{b}_{0}^\dagger+\hat{b}_{0}\hat{b}_{M_\mathrm{S}}^\dagger\right)\right\}\exp\left\{i 2 \tilde{d}   \lambda_1 D_\mathrm{F}^{\frac{1}{4}} \hat{b}_{0}^\dagger\hat{b}_{0}\right\}.
\end{aligned}
\end{equation}
The Baker–Campbell–Hausdorff (BCH) formula~\cite{magnus1954exponentialexpansion} was used to expand the exponential term,
\begin{equation}
e^{t(\hat{\mathcal{A}}+\hat{\mathcal{B}})}=e^{t \hat{\mathcal{A}}} e^{t \hat{\mathcal{B}}} e^{-\frac{t^2}{2}[\hat{\mathcal{A}}, \hat{\mathcal{B}}]} e^{\frac{t^3}{6}(2[\hat{\mathcal{B}},[\hat{\mathcal{A}}, \hat{\mathcal{B}}]]+[\hat{\mathcal{A}},[\hat{\mathcal{A}}, \hat{\mathcal{B}}]])}\cdots
\end{equation}
where $[\hat{\mathcal{A}},\hat{\mathcal{B}}]=\hat{\mathcal{A}}\hat{\mathcal{B}}-
\hat{\mathcal{B}}\hat{\mathcal{A}}$ is the commutator. In the limit of $2\tilde{d}D_\mathrm{F}^\frac{1}{4}\ll 1$, $\hat{U}_2(\tilde{d})$ is well approximated by a single-mode phase transformation followed by a beam splitter transformation. 
As a next step, we show the unitary matrix representation of the phase and beam splitter transformations. We keep track of both $\hat{b}_0$ and $\hat{b}_{M_\mathrm{S}}$ modes. The unitary matrices representing the phase transformation and the beam splitter transformation are~\cite{ferraro2005gaussianstatereview},
\begin{equation}
    \mathcal{M}_1=\left(\begin{array}{cc}
e^{2i\tilde{d}\lambda_1D_\mathrm{F}^\frac{1}{4} }  & 0 \\
        0 & 1
\end{array}\right),\quad
    \mathcal{M}_2=\left(\begin{array}{cc}
        \cos (2\tilde{d}D_\mathrm{F}^\frac{1}{4}) & i\sin (2\tilde{d}D_\mathrm{F}^\frac{1}{4})   \\
    i\sin (2\tilde{d}D_\mathrm{F}^\frac{1}{4})     &\cos (2\tilde{d}D_\mathrm{F}^\frac{1}{4})
    \end{array}\right).
\end{equation}
The overall parameter encoding unitary matrix is $\mathcal{M}_2\mathcal{M}_1$, which can be used to transform field operators in the Heisenberg picture. For a unitary matrix, $\mathcal{M}_2\mathcal{M}_1$, in order to transform the mean vector and covariance matrix, we can use the corresponding \textit{symplectic matrix}~\cite{weedbrook2012gaussian},
\begin{equation}
    \mathcal{S}=\left(\begin{array}{cc}
        \mathbbm{Re}[\mathcal{M}_2\mathcal{M}_1]&-\mathbbm{Im}[\mathcal{M}_2\mathcal{M}_1]\\
       \mathbbm{Im}[\mathcal{M}_2\mathcal{M}_1]  &   \mathbbm{Re}[\mathcal{M}_2\mathcal{M}_1] \end{array}\right),
       \label{eq:unitosym}
\end{equation}
where $\mathbbm{Re}$ returns the real part of the input and $\mathbbm{Im}$ returns the imaginary part of the input. The mean vector and covariance matrix are transformed into $X^\prime=\mathcal{S}X$, and $V^\prime=\mathcal{S}V\mathcal{S}^T$. 

The receiver shown in Fig.~\ref{fig:3}(a) is a two-mode homodyne measurement, which first applies a phase rotation $\nu_0=\mathrm{ArcTan}(e^{2r})$ to mode $\hat{b}_0^\prime$ and a phase rotation $\nu_1=\frac{\pi}{4}$ to mode $\hat{b}_{M_\mathrm{S}}^\prime$. Then homodyne detections are performed along the real quadratures of the two-mode state. The unitary matrix representing the phase evolution is, 
\begin{equation}
\mathcal{M}_3=\left(\begin{array}{cc}
e^{i \mathrm{ArcTan}(e^{2r})}&0\\
0&e^{i\frac{\pi}{4}}
\end{array}\right).
\end{equation}
We can write out the corresponding symplectic matrix, $\mathcal{S}_3$, by substituting $\mathcal{M}_2\mathcal{M}_1$ with $\mathcal{M}_3$ in Eq.~\ref{eq:unitosym}. The mean vector and covariance matrix after this phase rotation is $X^{\prime\prime}=\mathcal{S}_3X^\prime$, and $V^{\prime\prime}=\mathcal{S}_3V^\prime\mathcal{S}_3^T$.

The two-mode homodyne measurement on each temporal mode---among the $M_\mathrm{F} \sim WT$ orthonormal temporal-spectral modes contained in the probe duration $T$ and the probe's optical bandwidth $W$---along the real quadratures of the received state produces a measurement result, which are the realizations of two random variables following the multi-variate normal distribution $\{\mathbb{Q}_1,\mathbb{Q}_2\}^T\sim \mathcal{N}_2(\overline{X}, \Sigma)$, where the mean vector and covariance matrices are,
\begin{equation}\begin{aligned}
    \overline{X}=&\{X^{\prime\prime}_{\{1\}},X^{\prime\prime}_{\{2\}}\}^T=\{0,0\}^T,\\
    \Sigma=&\left(\begin{array}{cc}
    V^{\prime\prime}_{\{1,1\}} & V^{\prime\prime}_{\{1,2\}} \\
 V^{\prime\prime}_{\{2,1\}}   & V^{\prime\prime}_{\{2,2\}}
\end{array}\right).
\end{aligned}
\end{equation}
The elements in the covariance matrix $\Sigma$ are
\begin{equation}
    \begin{aligned}
V_{\{1,1\}}^{\prime\prime}  &  = \sin^2 \left(2 \tilde{d} D_\mathrm{F}^{\frac{1}{4}}\right)+e^{-2 r} \cos^2\left(2 \tilde{d} D_\mathrm{F}^{\frac{1}{4}}\right)\left(e^{4 r} \cos^2\left(\nu_0+2 \tilde{d}D_\mathrm{F}^{\frac{1}{4}} \lambda_1\right)+\sin^2\left(\nu_0+2 \tilde{d} D_\mathrm{F}^{\frac{1}{4}} \lambda_1\right)\right), \\
 V_{\{1,2\}}^{\prime\prime}&=\frac{1}{2} \sin\left(4 \tilde{d} D_\mathrm{F}^{\frac{1}{4}}\right)\left(2 \sin\left(\nu_0-\nu_1\right) \sinh(r)^2-\sin\left(\nu_0+\nu_1+4 \tilde{d} D_\mathrm{F}^{\frac{1}{4}} \lambda_1\right) \sinh(2 r)\right),\\
 V_{\{2,1\}}^{\prime\prime}&= V_{\{1,2\}}^{\prime\prime},\\
  V_{\{2,2\}}^{\prime\prime}&=\cos^2\left(2 \tilde{d} D_\mathrm{F}^{\frac{1}{4}}\right)+\sin^2\left(2 \tilde{d} D_F^{\frac{1}{4}}\right)\left(\cosh(2 r)-\cos\left(2\left(\nu_1+2 \tilde{d} D_\mathrm{F}^{\frac{1}{4}} \lambda_1\right)\right) \sinh(2 r)\right).
    \end{aligned}
\end{equation}
The angles are set to specific values, $\nu_0=\mathrm{ArcTan}(e^{2r})$ and $\nu_1=\frac{\pi}{4}$. The probability distribution is, 
\begin{equation}
    P(\mathbbm{q}_1,\mathbbm{q}_2)= \frac{1}{2\pi \mathrm{det}(\Sigma)^{1/2}} \exp{-\frac{1}{2} \left(\{\mathbbm{q}_1,\mathbbm{q}_2\}-\overline{X}\right) \Sigma^{-1}\left(\{\mathbbm{q}_1,\mathbbm{q}_2\}-\overline{X}\right)^T}.
\end{equation}
We use $\mathrm{det}(\Sigma)$ to denote the determinant of $\Sigma$, and $\Sigma^{-1}$ to denote the inverse of $\Sigma$.

The classical Fisher information is related to this probability distribution,
\begin{equation}
    J=\iint\displaylimits_{-\infty}^{\infty}\left(\frac{\partial P(\mathbbm{q}_1,\mathbbm{q}_2)}{\partial \tilde{d}}\frac{1}{P(\mathbbm{q}_1,\mathbbm{q}_2)}\right)^2 P (\mathbbm{q}_1,\mathbbm{q}_2)\mathrm{d}\mathbbm{q}_1\mathrm{d}\mathbbm{q}_2.
\end{equation}
Carrying out this integral, we verified that the CFI for this measurement saturates the QFI given in Eq.~\eqref{eq:b1qfi}. This implies that in the high-$M_\mathrm{F}$ limit (long integration time), the CFI of the two-mode homodyne detection receiver achieves the QFI afforded by the single-mode squeezed-vacuum-state probe of mode $\hat{b}_0$.

\section{Single-mode lossy parameter encoding channel}\label{app:lossy_channel}
Starting with an arbitrary linear superposition of $M_\mathrm{S}$ HG spatial modes at the transmitter, we represent the diffraction loss in terms of mode coefficients. Then we construct the \textit{principal component} basis for an arbitrary spatial mode displaced along $x$-axis, and present the corresponding Hamiltonian~\cite{2023mode_parameter_estimation}. 

A most general single-mode pure Gaussian state is a displaced squeezed state~\cite{weedbrook2012gaussian}, $\ket{\chi}\equiv\ket{\boldsymbol{\alpha};r e^{i \vartheta_1}}$, where $\boldsymbol{\alpha} \in {\mathbb C}$, $0\leq\vartheta< \pi$ and $r\ge0$. We allow the excitation of an arbitrary spatial mode $\Psi_0(x,y)$---living in the modal support of the first $M_\mathrm{S}$ HG spatial mode at the transmitter---in $\ket{\chi}$, 
\begin{equation}
\Psi_0(x,y)=\sum_{n=0}^{M_\mathrm{S}-1} c_n e^{i \theta_n} \Phi_{n,0}(x,y).
\end{equation}
This spatial mode is normalized, $\sum\displaylimits_{n=0}^{M_\mathrm{S}-1} c_n^2=1$.  
The transmissivity for this mode propagating through a free-space channel enclosed by soft-Gaussian apertures is $\eta_{\psi}=\sum\displaylimits_{n=0} c_n^2 \eta^{(n+1)}$. At the receiver, when there is no displacement, $d=0$, the excited mode is,
\begin{equation}
    \psi_0(x,y)=\frac{1}{\sqrt{\eta_{\psi}}}\sum_{n=0}^{M_\mathrm{S}-1} c_n e^{i \theta_n} \eta^{(n+1)/2}\phi_{n,0}(x,y).
\end{equation}
When this mode is displaced transversely along $x$-axis at the receiver aperture,  we perform a Taylor series expansion with respect to $d$ on the displaced mode,
\begin{equation}
\begin{aligned}\label{eq:PCA_expansion}
 \frac{\psi_0(x-d,y)}{\mathcal{A}_{\mathrm{R}}(x-d,y)} \mathcal{A}_{\mathrm{R}}(x,y)=&\psi_0(x-d,y) \exp\left\{\frac{(x-d)^2+y^2}{r_\mathrm{R}^2}\right\}
 \exp\left\{-\frac{x^2+y^2}{r_\mathrm{R}^2}\right\},\\
 =& \psi_0(x-d,y) \exp\left\{\frac{-2xd+d^2}{r_\mathrm{R}^2}\right\},\\
   \approx & \left(\psi_0(x,y)-d \frac{\partial \psi_0(x,y)}{\partial x}\right)\left(1+\frac{(-2xd+d^2)}{r_\mathrm{R}^2}\right),\\
    \approx & \psi_0(x,y)-\epsilon \frac{d}{r_\mathrm{R}} \psi_1^\mathrm{aux}(x,y)=\psi_0(x,y)-\epsilon \tilde{d} \psi_1^\mathrm{aux}(x,y),\\
\end{aligned}
\end{equation}
where $\frac{d}{r_\mathrm{R}}=\tilde{d}$, $\mathcal{A}_\mathrm{R}(x,y)$ is the receiver plane amplitude transmittance function and 
\begin{equation}
\begin{aligned}
    \psi_1^\mathrm{aux}(x,y)&\equiv\frac{r_R}{\epsilon} \left(\frac{\partial \psi_0(x,y)}{\partial x}+\frac{2 x}{r_R^2} \psi_0(x,y)\right),\\
         \epsilon&\equiv\left[ r_R^2\iint ^{+\infty}_{-\infty} \mathrm{d}x\mathrm{d}y \left(\frac{\partial \psi_0(x,y)}{\partial x}+\frac{2 x}{r_R^2} \psi_0(x,y)\right)\left(\frac{\partial \psi_0(x,y)^*}{\partial x}+\frac{2 x}{r_R^2} \psi_0(x,y)^*\right)\right]^{1/2}.
\end{aligned}
\end{equation}
We use $\epsilon$ to denote the total evolution strength, where $\epsilon\in \mathbb{R}$. We use $\psi_1^\mathrm{aux}(x,y)$ to denote a normalized mode. Applying Gram-Schmidt process to $\{\psi_0,\psi_1^\mathrm{aux}\}$ yields a mode $\psi_1$ which is orthogonal to $\psi_0$. 
\begin{equation}
\begin{aligned}
    \epsilon_0 \equiv & \iint \displaylimits^{+\infty}_{-\infty}   \epsilon\psi_1^\mathrm{aux}(x,y)   \psi_0^{\ast}(x,y) \mathrm{d}x\mathrm{d}y.\\
\epsilon_1 \psi_1
\equiv &\epsilon \psi_1^\mathrm{aux}(x,y) - \epsilon_0 \psi_0(x,y).
\end{aligned}\label{eq:epsilon_def}
\end{equation}
where the total evolution strength, $\epsilon$ is divided into two parts, $\{\epsilon_0,\epsilon_1\}$, in which $\epsilon_0\in\mathbb{C}$ and $\epsilon_1\in\mathbb{R}$. Invoking mode normalization, we have $\epsilon_1^2+|\epsilon_0|^2=\epsilon^2$.
Hence, we can write Eq.~\eqref{eq:PCA_expansion} in the form,
\begin{equation}
   \psi_0(x,y)-\epsilon \tilde{d} \psi_1^\mathrm{aux}(x,y)
  = \psi_0(x,y)-\tilde{d}  \epsilon_0  \psi_0(x,y)-\tilde{d} \epsilon_1 \psi_1(x,y).
\end{equation}
We study this in the regime $D_\mathrm{F}\gg1$, where the crosstalk matrix $Q_1(\tilde{d})$ is a skew Hermitian matrix. Thus given the modal decomposition above, the corresponding Hamiltonian is~\cite{2023mode_parameter_estimation},
\begin{equation}
    \hat{H}_3=-i \epsilon_0 \hat{\gamma}_0^\dagger \hat{\gamma}_0+i \epsilon_1(\hat{\gamma}_0^\dagger \hat{\gamma}_1-\hat{\gamma}_1^\dagger \hat{\gamma}_0).
\end{equation}
in which $\{\hat{\gamma}_0,\hat{\gamma}_1\}$ represent the field operators in mode $\{\psi_0,\psi_1\}$ respectively. The first term describes the evolution of the initially excited mode, and the second term describes the interaction between the excited mode with a mode starting in a vacuum state. In the limit of small $\tilde{d}$, the unitary transformation can be expanded using BCH as
    \begin{equation}
    \begin{aligned}
     \hat{U}_3(\tilde{d})\equiv&\exp{-i\tilde{d}\hat{H}_3}
        =\exp{-\tilde{d}\epsilon_0 \hat{\gamma}_0^\dagger \hat{\gamma}_0+\tilde{d} \epsilon_1 (\hat{\gamma}_0^\dagger \hat{\gamma}_1-\hat{\gamma}_1^\dagger \hat{\gamma}_0)},\\
        =& e^{\tilde{d} \epsilon_1 (\hat{\gamma}_0^\dagger \hat{\gamma}_1-\hat{\gamma}_1^\dagger \hat{\gamma}_0)}
        e^{-\tilde{d}\epsilon_0 \hat{\gamma}_0^\dagger \hat{\gamma}_0}
        e^{\frac{-\tilde{d}^2}{2} \left[\epsilon_1 (\hat{\gamma}_0^\dagger \hat{\gamma}_1-\hat{\gamma}_1^\dagger \hat{\gamma}_0),-\epsilon_0 \hat{\gamma}_0^\dagger \hat{\gamma}_0\right]}e^{\frac{\tilde{d}^3}{6} \dots}\cdots,\\
        \approx& e^{\tilde{d} \epsilon_1 (\hat{\gamma}_0^\dagger \hat{\gamma}_1-\hat{\gamma}_1^\dagger \hat{\gamma}_0)}
        e^{-\tilde{d}\epsilon_0 \hat{\gamma}_0^\dagger \hat{\gamma}_0},
    \end{aligned}
\end{equation}
where the square bracket represents a commutator and the omitted terms only depend on $\tilde{d}$ to higher orders \cite{magnus1954exponentialexpansion}. The transformation induced by a transverse displacement can be approximated by a phase accrued on the excited mode, $\psi_0$, and a beam splitter transformation between the excited mode $\psi_0$ and mode $\psi_1$ which is in a vacuum state. We show the quantum circuit representation of $\hat{U}_3(\tilde{d})$ in Fig.~\ref{fig:4}(a).

For an arbitrary mode $\Psi_0$, in terms of the mode coefficients and the diffraction loss, the evolution strengths are: 
\begin{equation}
\begin{aligned}
    \epsilon^2
 &=4\sqrt{ D_\mathrm{F}} \left\{ 1+\frac{2}{\eta_\psi} \left[\sum\displaylimits_{n=0}^{M_\mathrm{S}-1} c_n^2 n \eta^{n+1}+\sum\displaylimits_{n=0}^{M_\mathrm{S}-1} c_n c_{n+2}\sqrt{(n+1)(n+2)} \eta^{n+2} \cos{\left(\theta_n-\theta_{n+2}+\frac{\pi}{2}\right)}\right]\right\},\\
     \epsilon_0    &=4\frac{i ( D_\mathrm{F})^{\frac{1}{4}}}{\eta_\psi}\left\{\sum\displaylimits_{n=0}^{M_\mathrm{S}-1} c_n c_{n+1} \eta^{(2n+3)/2} \sqrt{n+1}  \cos{\left(\theta_n-\theta_{n+1}+ \frac{\pi}{4}\right)}\right\},
\label{eq:mode_parameter}
    \end{aligned}
\end{equation}
and $\epsilon_1$ can be obtained through $\epsilon_1^2=\epsilon^2-|\epsilon_0|^2$. We use the crosstalk matrix in the large $D_\mathrm{F}$ limit, $Q_1(\tilde{d})$ to obtain $\psi_1^\mathrm{aux}(x,y)$ represented in HG basis, then arrive at the evolution strengths above.

\section{QFI of a phase parameter encoded on a single-mode Gaussian state}\label{app:lossyQFI}
In this section, we use the mean vector and covariance matrix representation of a Gaussian state and the symplectic matrix representation of a unitary transformation~\cite{weedbrook2012gaussian} to find and evaluate $K\left[\hat{\rho}(0), -i\epsilon_0\hat{N}_0\right]$~\cite{pinel2013quantumsinglemodegaussian_single_parameter}, which is the first term in Eq~\eqref{eq:qfi_mode_and_state}. First we write out the mean vector and covariance matrix representation of a most general single-mode Gaussian state. Then we show how they are transformed by the lossy channel and the parameter encoding process. With the parameter carrying state, we evaluate $K\left[\hat{\rho}(0), -i\epsilon_0\hat{N}_0\right]$. 

At the transmitter, we have a single-mode displaced squeezed state, $\ket{\chi}=\ket{\boldsymbol{\alpha};r e^{i\vartheta_1}}$, where $\boldsymbol{\alpha}= \alpha e^{i \vartheta_0}$, $\alpha\in\mathbb{R}$, $0\leq\vartheta_0<2\pi$, $0\leq\vartheta_1<\pi$ and $r>0$. We include $\vartheta_1$ to denote the squeezing angle. In terms of the displacement operator and squeezing operator, the displaced squeezed state is defined as
\begin{equation}
    \ket{\boldsymbol{\alpha};r e^{i\vartheta_1}}\equiv\hat{
\mathcal{R}}(\vartheta_0)\hat{D}(\boldsymbol{\alpha})\hat{S}(r e^{i \vartheta_1})\ket{0}.
\end{equation}
We can use $\hat{\beta}_0^{}$ to denote the field operator corresponding to mode $\Psi_0$. The single mode phase rotation operator is $\hat{\mathcal{R}}(\vartheta_0)\equiv\exp(-i\vartheta_0\hat{\beta}_0^{\dagger}\hat{\beta}_0^{})$. The displacement operator is $\hat{D}\left(\boldsymbol{\alpha}\right)\equiv\exp \left\{\boldsymbol{\alpha} \hat{\beta}_0^{\dagger}-\boldsymbol{\alpha}^* \hat{\beta}_0^{}\right\}$. And the squeezing operator is $\hat{S}(r e^{i \vartheta_1})\equiv\exp \left\{\frac{1}{2} r e^{i \vartheta_1}\left(\hat{\beta}_0^{\dagger}\right)^2-\frac{1}{2} r e^{- i \vartheta_1} \left(\hat{\beta}_0^{}\right)^2\right\}$~\cite{ferraro2005gaussianstatereview,weedbrook2012gaussian}. For the analysis of diffraction loss we include an environmental vacuum mode. Thus at the transmitter, the mean vector and covariance matrix of the state is~\cite{weedbrook2012gaussian},
\begin{equation}
  X_0=\{2\alpha \cos(\vartheta_0),0,-2\alpha \sin(\vartheta_0),0\}^T,
\end{equation}
\begin{equation}
  V_0=\left(\begin{array}{cccc}
\cosh(2 r)+\cos(\vartheta_1+2\vartheta_0) \sinh(2 r)& 0& -\sin(\vartheta_1+2\vartheta_0) \sinh(2 r) &0\\
0&1&0&0\\
-\sin(\vartheta_1+2\vartheta_0) \sinh(2 r) &0& \cosh(2 r)-\cos(\vartheta_1+2\vartheta_0) \sinh(2 r)&0\\
0&0&0&1\\
\end{array}\right). 
\end{equation}
For this state, the total mean photon number is $\alpha^2+\mathrm{sinh}^2 r=N$. The free space propagation is represented as a beam splitter interaction of transmissivity $\eta_\psi$ between the excited mode and an environmental mode in a vacuum state. We can use an unitary matrix to represent this beam splitter,
\begin{equation}
    \mathcal{M}_\mathrm{BS}=\left(\begin{array}{cc}
\sqrt{\eta_\psi }        & i \sqrt{1-\eta_\psi }   \\
   i \sqrt{1-\eta_\psi }       & \sqrt{\eta_\psi }  
    \end{array}\right).
\end{equation}
The corresponding symplectic matrix, $\mathcal{S}_\mathrm{BS}$, is obtained by applying Eq.~\eqref{eq:unitosym}. The transformed mean vector and covariance matrix is $X_1=\mathcal{S}_\mathrm{BS}X_0$ and $V_1=\mathcal{S}_\mathrm{BS}V_0\mathcal{S}_\mathrm{BS}^{T}$. At the receiver, tracing out the environmental mode, we have access to a single-mode state with mean vector and covariance matrix, 
\begin{equation}
 X_2=\sqrt{\eta_\psi}\{2\alpha \cos(\vartheta_0),-2\alpha \sin(\vartheta_0)\}^T,
\end{equation}
\begin{equation}
   V_2=\left(\begin{array}{cc}
1-\eta_{\psi}+\eta_{\psi}\left(\cosh\left(2 r\right)+\cos\left(\vartheta_1+2\vartheta_0\right) \sinh\left(2r\right)\right) & -\eta_{\psi} \sin\left(\vartheta_1+2\vartheta_0\right) \sinh\left(2 r\right) \\
-\eta_{\psi} \sin\left(\vartheta_1+2\vartheta_0\right) \sinh\left(2 r\right) & 1-\eta_{\psi}+\eta_{\psi}\left(\cosh\left(2 r\right)-\cos\left(\vartheta_1+2\vartheta_0\right) \sinh\left(2 r\right)\right)
\end{array}\right).
\end{equation}

In order to apply the result from~\cite{pinel2013quantumsinglemodegaussian_single_parameter}, we first find the symplectic form of $V_2$~\cite{weedbrook2012gaussian}, 
\begin{equation}
\begin{aligned}
   V_2= &\left(\begin{array}{cc}
\cos(\vartheta_0) & \sin(\vartheta_0) \\
-\sin(\vartheta_0) & \cos(\vartheta_0) 
\end{array}\right)\left(\begin{array}{cc}
\cos(\frac{\vartheta_1}{2}) & \sin(\frac{\vartheta_1}{2}) \\
-\sin(\frac{\vartheta_1}{2}) & \cos(\frac{\vartheta_1}{2}) 
\end{array}\right) \left(\begin{array}{cc}
1+\left(-1+e^{2 r}\right) \eta_{\psi}& 0 \\
0 & 1+\left(-1+e^{-2 r}\right) \eta_\psi
\end{array}\right)\\
&\left(\begin{array}{cc}
\cos(\frac{\vartheta_1}{2}) & -\sin(\frac{\vartheta_1}{2}) \\
\sin(\frac{\vartheta_1}{2}) & \cos(\frac{\vartheta_1}{2}) 
\end{array}\right)\left(\begin{array}{cc}
\cos(\vartheta_0) & -\sin(\vartheta_0) \\
\sin(\vartheta_0) & \cos(\vartheta_0) 
\end{array}\right),\\
=&\left(\begin{array}{cc}
\cos(\vartheta_0) & \sin(\vartheta_0) \\
-\sin(\vartheta_0) & \cos(\vartheta_0) 
\end{array}\right)\left(\begin{array}{cc}
\cos(\frac{\vartheta_1}{2}) & \sin(\frac{\vartheta_1}{2}) \\
-\sin(\frac{\vartheta_1}{2}) & \cos(\frac{\vartheta_1}{2}) 
\end{array}\right) 
\left(\begin{array}{cc}
e^{r^\prime}& 0 \\
0 & e^{-r^\prime}
\end{array}\right)
\left(\begin{array}{cc}
v& 0 \\
0 & v
\end{array}\right)
\left(\begin{array}{cc}
e^{r^\prime}& 0 \\
0 & e^{-r^\prime}
\end{array}\right)\\
&\left(\begin{array}{cc}
\cos(\frac{\vartheta_1}{2}) & -\sin(\frac{\vartheta_1}{2}) \\
\sin(\frac{\vartheta_1}{2}) & \cos(\frac{\vartheta_1}{2}) 
\end{array}\right)\left(\begin{array}{cc}
\cos(\vartheta_0) &- \sin(\vartheta_0) \\
\sin(\vartheta_0) & \cos(\vartheta_0) 
\end{array}\right),
\end{aligned}
\end{equation}
where $v$ represents the state purity~\cite{weedbrook2012gaussian}, and $r^\prime$ represents the effective squeezing parameter. 
\begin{equation}\label{eq:vandrprime}
    \begin{aligned}
        r^\prime&\equiv\frac{1}{4} \mathrm{ln}\left( \frac{1-\eta_\psi+ e^{2r} \eta_\psi }{ 1-\eta_\psi+ e^{-2r} \eta_\psi}\right),\\ 
v&\equiv\sqrt{( 1-\eta_\psi+ e^{-2r} \eta_\psi)( 1-\eta_\psi+ e^{2r} \eta_\psi)}.\\
    \end{aligned}
\end{equation}
After propagating through the lossy channel, the state becomes mixed with $v>1$, and the effective squeezing will be less than the initial squeezing due to loss. For a single mode phase evolution with strength $\epsilon_0$, the corresponding symplectic matrix is,
\begin{equation}
    \mathcal{S}_\mathrm{d}=\left(\begin{array}{cc}
       \cos(|\epsilon_0|\tilde{d})  &   \sin(|\epsilon_0|\tilde{d})   \\
           -\sin(|\epsilon_0|\tilde{d})  &    \cos(|\epsilon_0|\tilde{d}) 
    \end{array}\right).
\end{equation}
The parameter carrying state's mean vector and covariance matrix is $X_3= \mathcal{S}_\mathrm{d} X_2$ and $V_3= \mathcal{S}_\mathrm{d} V_2 \mathcal{S}_\mathrm{d}^T$.

Applying Eq. (12) from ~\cite{pinel2013quantumsinglemodegaussian_single_parameter}, we have,
\begin{equation}
    K\left[\hat{\rho}(0),-i\epsilon_0 \hat{N}_0\right]=\frac{1}{2} \frac{\tr\left[\left(V_3^{-1} V_3^{\prime}\right)^2\right]}{1+1/v^2}+2 \frac{(1/v)^{\prime 2}}{1-1/v^4}+\Delta X_3^{\prime \top} V_3^{-1} \Delta X_3^{\prime},
\end{equation}
where $\Delta X_3^\prime=\frac{\mathrm{d} \left[X_3(\tilde{d}+\varepsilon)-X_3(\tilde{d})\right]}{\mathrm{d}\varepsilon}|_{\varepsilon\rightarrow0}$ and $V_3^\prime$ represent the element wise derivative of $V_3$ with respect to $\tilde{d}$. Evaluating this expression we get,
\begin{equation}
\begin{aligned}\label{eq:QFI_singlemode_phase}
K\left[\hat{\rho}(0),-i\epsilon_0 \hat{N}_0\right]&=\frac{4 |\epsilon_0|^2\mathrm{sinh}^2(2r^\prime)}{1+v^{-2}}+\frac{4|\epsilon_0|^2 \alpha^2}{v} \eta_\psi\left(\cosh\left(2r^\prime\right)+\cos\left(\theta_1\right)\sinh(2r^\prime)\right).
\end{aligned}
\end{equation}
By taking $\vartheta_1=0$, the above expression is maximized,
\begin{equation}\label{eq:testing1}
K\left[\hat{\rho}(0),-i\epsilon_0 \hat{N}_0\right]=\frac{4 |\epsilon_0|^2\mathrm{sinh}^2(2r^\prime)}{1+v^{-2}}+\frac{4|\epsilon_0|^2 \alpha^2}{v} \eta_\psi\exp(2r^\prime).
\end{equation}

\section{QFI optimization merit function simplification}\label{app:simplification of merit function}
Before performing optimization, we simplify the merit function to reduce degeneracy of the parameter space to ensure a robust optimization.

Rearranging the terms within Eq.~\eqref{eq:qfi_mode_and_state} to separate the terms depending on $|\epsilon_0|^2$ and $\epsilon^2$, yielding
\begin{equation}
K\left[\hat{\tau}(0),\hat{H}_3\right]=
|\epsilon_0|^2\left(K\left[\hat{\rho}(0), -\hat{N}_0\right]-4\eta_\psi N\right)+4\epsilon^2\eta_\psi N.
\label{eq:testing}
\end{equation}
It is easy to show that with optimized power allocation between squeezing and displacement for different levels of loss, $\eta_\psi$, we have $K\left[\hat{\rho}(0), -i\hat{N}_0\right]\geq 4\eta_\psi N$. This verifies that a general single-mode state will at least provide the sensitivity attainable by a coherent state for phase estimation. Thus we seek to maximize both $|\epsilon_0|^2$ and $\epsilon_1^2$. Note that the phase values maximizing $|\epsilon_0|^2$ require, $\theta_{n+1}=\theta_n+\pi/4$, and the phase values maximizing $\epsilon^2$ require, $\theta_{n+2}=\theta_n+\pi/2$, per inspection of Eq.~\eqref{eq:mode_parameter}. The first condition is sufficient but unnecessary for the second condition to hold. We can simplify Eq.~\eqref{eq:mode_parameter} with these conditions. 

We summarize the mode-depend parameters $\{\epsilon_0,\epsilon,\eta_\psi\}$, 
\begin{equation}
\begin{aligned}
    \epsilon^2
 &=\frac{2\sqrt{4 D_\mathrm{F}}}{r_R^2} \left\{ 1+\frac{2}{\eta_\psi} \left[\sum\displaylimits_{n=0}^{M_\mathrm{S}-1} c_n^2 n \eta^{n+1}+\sum\displaylimits_{n=0} c_n c_{n+2}\sqrt{(n+1)(n+2)} \eta^{n+2} \right]\right\},\\
    \epsilon_0&=\frac{2}{r_R}\frac{i (16 D_\mathrm{F})^{\frac{1}{4}}}{\eta_\psi}\left\{\sum\displaylimits_{n=0}^{M_\mathrm{S}-1} c_n c_{n+1} \eta^{(2n+3)/2} \sqrt{n+1} \right\},\\
\eta_{\psi}&=\sum\displaylimits_{n=0}^{M_\mathrm{S}-1} c_n^2 \eta^{(n+1)}.
\end{aligned}  \label{eq:mode_parameter_simplified_3}
\end{equation}

We summarize the state-dependent parameters, 
\begin{equation}    \label{eq:state_function_summary}
    \begin{aligned}
& r^\prime\equiv\frac{1}{4} \mathrm{ln}\left( \frac{1-\eta_\psi+ e^{2r} \eta_\psi }{ 1-\eta_\psi+ e^{-2r} \eta_\psi}\right),\\ 
&v^{}_{}\equiv\sqrt{( 1-\eta_\psi+ e^{-2r} \eta_\psi)( 1-\eta_\psi+ e^{2r} \eta_\psi)},\\
&N=\sinh(r)^2+\alpha^2,\\
&N_1=\alpha^2, N_2=\sinh(r)^2.\\
\end{aligned}
\end{equation}

Then the QFI functions are
\begin{equation}    \label{eq:qfi_summ}
    \begin{aligned}
&K\left[\hat{\tau}(0),\hat{H}_3\right]=
K\left[\hat{\rho}(0), -i\epsilon_0\hat{N}_0\right]+4\left[\epsilon^2-|\epsilon_0|^2\right]\langle\hat{N}_0\rangle_{\hat{\rho}(0)},\\
&K\left[\hat{\rho}(0), -\hat{N}_0\right]=\frac{4 \sinh^2(2r^\prime)}{1+v^{-2}}+\frac{4 \alpha^2}{v} \eta_\psi\exp(2r^\prime),\\
&\langle\hat{N}_0\rangle_{\hat{\rho}}=N \eta_{\psi}.
\end{aligned}
\end{equation}
Equipped with Eq.~\eqref{eq:mode_parameter_simplified_3},
Eq.~\eqref{eq:state_function_summary} and Eq.~\eqref{eq:qfi_summ}, we perform optimization to find the best power allocation, $\{N_1,N_2\}$, and the mode coefficients $\{c_n,\theta_n\}$ given certain combinations of resources, $\{N, M_\mathrm{S}, D_\mathrm{F}\}$. The optimized results are shown in Fig.~\ref{fig:4} and Fig.~\ref{fig:5}. 

\section{The effect of diffraction loss on optimal mode for squeezed-light probe}\label{app:optmodecn}

\begin{figure}
    \centering\includegraphics[width=\linewidth]{ 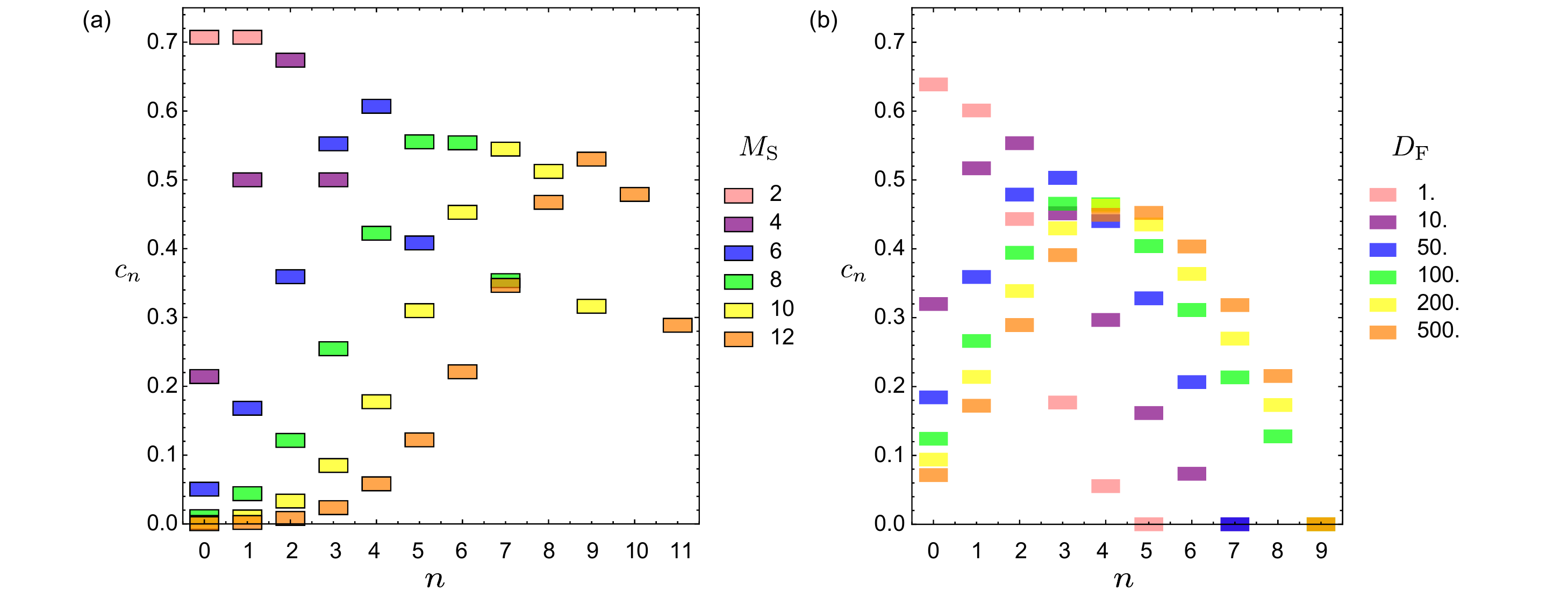}
    \caption{The amplitudes of the mode coefficients $c_n$, in the transmitter-side HG basis, of: (a) the $\hat{b}_0$ mode (optimal mode to excite a squeezed-vacuum probe in, when the transmitter is restricted to employ a lossless $M_{\rm S}$-HG-mode support), with $M_\mathrm{S} = 2, 4, 6, 8, 10$ and $12$; and (b) the $Z_s$ mode (optimal mode to excite a squeezed-state probe in, for a given $D_{\rm F}$ but no $M_{\rm S}$ support) for $D_\mathrm{F}$ values ranging $1$ to $500$.}
    \label{fig:optmodecn}
\end{figure}
\end{widetext}
In Fig.~\ref{fig:3}(c) we show the spatial mode shapes of the $\hat{b}_0$ mode, which is the optimal mode to excite a squeezed-vacuum probe in, and one that attains HL sensitivity, conditioned on the transmitted probe light being artificially restricted to employ a lossless lowest-$M_{\rm S}$ HG modes' support with $M_{\rm S} \ll D_{\rm F}$, and the total probe energy being large. These modes have a `comet-like' shape, with the center of mass shifted laterally on one side of the origin and the shift being larger for larger $M_{\rm S}$. In contrast, in Fig.~\ref{fig:5}(b) we display the mode shapes of the $Z_s$ mode, which we find (numerically) is the QFI-optimal mode to excite, when the transmitter chooses to employ a single-mode displaced-squeezed-state probe. This optimization was performed subject to a given $D_{\rm F}$ but no restriction on the modal support, $M_{\rm S}$. The $Z_s$ modes visually appear similar to a laterally-shifted HG00 mode. 

Here we provide some additional numerical insights into the optimal mode shapes for these two scenarios: 
(a) lossless, with modal support restricted to $M_{\rm S}$, and (b) lossy, with diffraction-limited system geometry quantified by $D_{\rm F}$, and unrestricted modal support. See Fig.~\ref{fig:optmodecn}. We plot here the amplitudes $c_n$ of the coefficients of expansion in terms of $n$-th transmitter-side HG modes, $n=0, 1, 2, \ldots$ (referred to earlier as `mode coefficients'), of: (a) the $\hat{b}_0$ mode, with $M_\mathrm{S} = 2, 4, 6, 8, 10$ and $12$; and (b) the $Z_s$ mode for $D_\mathrm{F}$ values ranging $1$ (near-far-field boundary) to $500$ (deep near-field), respectively. We see in Fig.~\ref{fig:optmodecn}(a) that with $M_{\rm S}$ lossless modes, as $M_{\rm S}$ increases, the HG support of the optimal ($b_0$) mode increases rapidly. This is consistent both with the comet-like shape of the $b_0$ mode, and with the fact that high-order modes are preferred when loss is taken out of the equation. Fig.~\ref{fig:optmodecn}(a) shows that with $D_{\rm F}$ held fixed, the modal distribution of the QFI-optimal ($Z_s$) mode follows an approximately Poisson distribution in the centered HG basis, as expected for an off-center HG00 mode and is consistent with the numerically-optimized mode shape of the $Z_s$ modes shown in Fig.~\ref{fig:5}(b). Further, the shift to the right, of the center of mass of the $b_0$ modes, as seen in Fig.~\ref{fig:3}(c), as $M_{\rm S}$ increases---which is consistent with the $\left\{c_n\right\}$ distributions in Fig.~\ref{fig:optmodecn}(a)---is more rapid as compared to the shift to the right, of the center of mass of the $Z_s$ modes, as seen in Fig.~\ref{fig:5}(b), as $D_{\rm F}$ is increased, which is consistent with the $\left\{c_n\right\}$ distributions in Fig.~\ref{fig:optmodecn}(b). This is because the transmitter's propensity to using a high-order mode is higher if they are lossless.

\end{document}